\documentclass[twocolumn,tighten,twocolappendix]{aastex63}

\usepackage{CJK}
\graphicspath{{./}{figures/}}

\submitjournal{ApJ}

\shorttitle{Clustering of Stars in the Galactic Disk}
\shortauthors{Kamdar et al.}

\begin{document}
\begin{CJK*}{UTF8}{gbsn}

\title{Spatial and Kinematic Clustering of Stars in the Galactic Disk}

\correspondingauthor{Harshil Kamdar}
\email{harshil.kamdar@cfa.harvard.edu}

\author[0000-0001-5625-5342]{Harshil Kamdar}
\affiliation{Center for Astrophysics $|$ Harvard \& Smithsonian, 60 Garden Street, Cambridge, MA 02138, USA}

\author[0000-0002-1590-8551]{Charlie Conroy}
\affiliation{Center for Astrophysics $|$ Harvard \& Smithsonian, 60 Garden Street, Cambridge, MA 02138, USA}

\author[0000-0001-5082-9536]{Yuan-Sen Ting (丁源森)}
\affiliation{Institute for Advanced Study, Princeton, NJ 08540, USA}
\affiliation{Department of Astrophysical Sciences, Princeton University, Princeton, NJ 08544, USA}
\affiliation{Observatories of the Carnegie Institution of Washington, 813 Santa Barbara Street, Pasadena, CA 91101, USA}
\affiliation{Research School of Astronomy and Astrophysics, Mount Stromlo Observatory, Cotter Road, Weston Creek, ACT 2611, Canberra, Australia}

\author[0000-0002-6871-1752]{Kareem El-Badry}
\affiliation{Department of Astronomy and Theoretical Astrophysics Center, University of California Berkeley, Berkeley, CA 94720, USA}

\begin{abstract}
The Galactic disk is expected to be spatially, kinematically, and chemically clustered on many scales due to both star formation and non-axisymmetries in the Galactic potential. In this work we calculate the spatial and kinematic two-point correlation functions using a sample of $1.7 \times 10^6$ stars within 1 kpc of the Sun with 6D phase space information available from \textit{Gaia} DR2.  Clustering is detected on spatial scales of $1-300$ pc and velocity scales of at least 15 km s$^{-1}$. With bound structures included, the data have a power-law index ($\xi(\Delta r) \propto \Delta r^{\gamma}$) of $\gamma\approx-2$ at most spatial scales, which is in line with theoretical predictions. After removing bound structures, the data have a power-law index of $\gamma\approx-1$ for $1 < \Delta r < 100$ pc and $\gamma \lesssim -1.5$ for $>100$ pc. We interpret these results with the aid of a novel star-by-star simulation of the Galaxy in which stars are born in clusters orbiting in a realistic potential that includes spiral arms, a bar, and giant molecular clouds (GMCs). We find that the simulation largely agrees with the observations (within a factor of $2-3$) at all spatial and kinematic scales. In detail, the correlation function in the simulation is shallower than the data at $\lesssim20$ pc scales, and steeper than the data at $\gtrsim30$ pc scales.  We also find a persistent clustering signal in the kinematic correlation function for the data at large $\Delta v$ ($>5$ km s$^{-1}$) that is not present in the simulations. We speculate that this mismatch between observations and simulations may be due to two processes not included in the present simulation: hierarchical star formation and transient spiral arms. We also use these simulations to predict the clustering signal as a function of pair-wise metallicity and age separations. Ages and metallicities measured with a precision of $50\%$ and $0.05$ dex, respectively, are required in order to enhance the clustering signal beyond the current measurements.  
\end{abstract}


\section{Introduction} 
\label{sec:intro}

Much of our knowledge about how galaxies form and evolve in the Universe comes from detailed studies of our own Galaxy. The rich history of the Galaxy is encoded in the distribution of the kinematics and the chemistry of its stars. The unprecedented amount of astrometric \citep{brown2018gaia} and spectroscopic \citep[e.g.,][]{kollmeier2017sdss, kunder2017radial, buder2019galah, ahumada2019sixteenth} data expected in the coming years on our Galaxy will revolutionize our view of the different physical processes in galaxy evolution. To reconstruct the history of the Galaxy we must study the birth, evolution, and death of the building blocks of star formation -- star clusters. 

Most stars are thought to be born in a spatially and temporally correlated way \citep[see reviews by][]{bland2010long, krumholz2019star}; however, most stellar aggregates are quickly disrupted in the Galaxy \citep{lada2003embedded, gieles2006star}. Consequently, much effort has been devoted to the fossil record of star formation by inspecting the chemical make-up of stars to identify those that might have been born in the same birth cloud \citep[i.e., ``co-natal" stars;][]{freeman2002new, bland2010long, ting2015prospects,2020arXiv200404263P}. Recent work \citep[e.g.,][]{meingast2019extended, k19b, coronado2020birth} has also shown promise in using the kinematic properties of stars to find those that might have been born in the same cluster but have since drifted apart. 

Non-axisymmetric features in the Galactic disk\footnote{For the remainder of this paper, we use ``non-axisymmetries" to mean the bar, spiral arms, and GMCs.} can also have a significant impact on the structure in the disk \citep[e.g.,][]{hunt2018transient, trick2019galactic, sellwood2019discriminating}. Structure on large scales -- due to resonances in the kinematics and the enrichment history of the Galaxy in chemistry -- has been extensively studied in recent years in both the data \citep[e.g.,][]{kawata2018radial, michtchenko2018stellar, bland2019galah, trick2019galactic} and simulations \citep[e.g.,][]{fragkoudi2019ridges, monari2019signatures}.  However, there remain key questions about clustering on intermediate scales, where both correlated star formation and resonances are likely important. 

In \citet[][hereafter K19a]{k19a}, we presented a star-by-star dynamical model of the Galactic disk that takes into account the clustered nature of star formation and the complexity of the Galactic potential. The highest-resolution cosmological zoom-in simulations and isolated \textit{N}-body simulations of the Galaxy have stellar particles with masses $\gtrsim 500 M_{\odot}$, which is the typical mass of a star cluster born today.  These simulations are therefore unable to probe the scales relevant for studying the small-scale clustering of individual stars.  Our model self-consistently evolves 4 billion stars over the last 5 Gyr in a realistic time-varying potential that includes an axisymmetric component, a bar, spiral arms, and giant molecular clouds (GMCs). All stars are born in clusters with a subgrid model for cluster birth and dissolution \citep{lada2003embedded}. As direct \textit{N}-body calculations for billions of stars is computationally infeasible, we developed a method of initializing star clusters to mimic the effects of direct \textit{N}-body interactions. 

A key, unexpected prediction from K19a was that stars separated spatially by as much as $20-30$ pc but moving at similar relative velocities are likely co-natal. We used \textit{Gaia} DR2 and LAMOST DR4 data in \cite[][hereafter K19b]{k19b} to identify and study these ``co-moving" pairs of stars \citep[e.g.,][]{oh2017comoving} with both kinematic information and chemical abundances. In K19b, we identified 111 such co-moving pairs in the Solar neighborhood with reliable astrometric and spectroscopic measurements. These pairs showed a strong preference for having similar metallicities when compared to random field pairs, supporting the idea that they were born together. 

The co-moving pairs identified in K19b along with wide binaries from other work \citep[e.g.,][]{andrews2017wide2, el2018imprints, hawkins2020identical} probe (by design) fairly small spatial scales ($1-20$ pc). Other work studying resonances \citep[e.g.,][]{bovy2015power,khanna2019galah} probe structure at kiloparsec scales. Here we expand upon previous work by measuring the two-point correlation function for the solar neighborhood on physical scales from parsecs to kiloparsecs in order to study the clustered nature of the Galaxy.

The two-point correlation function (TPCF hereafter) has been extensively used in cosmology \citep[see references in][]{peebles2001galaxy} and other areas of physics \citep[e.g.,][]{kagan1980spatial, zamolodchikov1991two}. The TPCF characterizes the excess probability of two points separated by some $r$ relative to an unclustered distribution. Given its simplicity and relative ease of interpretability, the TPCF has been used to constrain cosmological models \citep[e.g.,][]{eisenstein2005detection, sanchez2012clustering, alam2017clustering}, study the galaxy-halo connection \citep[e.g.][]{wechsler2006dependence, conroy2006modeling, reddick2013connection}, probe the epoch of reionization \citep[e.g.,][]{mcquinn2007studying}, and quantify the clustering of young stellar clusters in other galaxies \citep[e.g.,][]{houlahan1990recognition, elmegreen2014hierarchical, grasha2017hierarchical, gouliermis2017hierarchical}. 

There has been some previous work characterizing the clustering in the Galactic halo \citep{cooper2011two, lancaster2019quantifying} to study kinematic substrutures and infer the Galactic accretion history. \cite{bovy2015power} and  \cite{khanna2019galah} calculated the power spectrum (the fourier tranform of the TPCF) of velocity fluctuations in the disk to study the dynamical influence of the bar in the disk. \cite{mao2015probing} used the TPCF to probe Galactic disk structure in SEGUE G-Dwarf stars. Mao et al. placed strong constraints on  of the scale heights of the thin disk and the thick disk. However, they also show the strong biases that the selection function and the non-uniform density profile of the Galaxy impart on the TPCF. Consequently, calculating the TPCF in Galactic science has been non-trivial up to now due to the relative dearth of data, the complex density profile of the Galaxy, and the absence of theory or simulations to guide predictions at all scales.  The landscape has changed dramatically with the release of \textit{Gaia} DR2, which provided 6D phase space information for millions of stars. Moreover, the star-by-star simulations presented in K19a enable, for the first time,  predictions of stellar clustering on both small and large spatial scales.

In this paper we present the spatial and velocity TPCF for stars in the solar neighborhood using \textit{Gaia} data and provide predictions from the simulations presented in K19a. The rest of this paper is organized as follows. Section \ref{sec:ds} discusses the quality cuts imposed on \textit{Gaia} data and the simulations in K19a; we also present the mock catalog from a smooth unclustered realization of the Galaxy \citep{rybizki2018gaia} as a control. In Section \ref{sec:tpcf} we introduce the TPCF, describe our method for the random catalog, and present several validation tests. The spatial and kinematic TPCFs for the data and the simulations are presented in Section \ref{sec:resultsp1}, and we predict the clustering in the disk when we combine kinematic data with metallicity and age information in Section \ref{sec:future}. A summary of our results is provided in Section \ref{sec:conclusions}. 

\section{Data \& Simulations}
\label{sec:ds}

\subsection{Observational Data}

We focus on stars with radial velocities in \textit{Gaia} DR2. We start with the 6D \textit{Gaia} DR2 \citep{brown2018gaia} catalog from \citet{marchetti2019gaia}. Stars were selected within a cylinder centered at $(X,Y,Z)=(-8.2, 0.0, 0.025)$ kpc \citep{bland2016galaxy} with a radius of $0.5$ kpc and a height of $1$ kpc (0.5 kpc above and below the solar position). The impact of the volume of the cylinder on the TPCF is discussed in Appendix A. We choose this volume to ensure that the data are of high quality, and because going out to a larger volume would require a careful treatment of the fluctuations in the local mid-plane of the Galaxy \citep{beane2019implications}. 

The distance to sources in the catalog with low relative error in parallax is calculated by simply inverting the parallax. For stars with a larger parallax uncertainty, the distances are calculated using the Bayesian approach outlined in \citet{bailer2018estimating}. The rotation velocity at the Sun's position is assumed to be $v_{LSR}=238$ km s$^{-1}$, and the Sun's orbital velocity is assumed to be $(U_{\odot}, V_{\odot}, W_{\odot})=(14.0,12.24,7.25)$ km s$^{-1}$ \citep{schonrich2012galactic, bland2016galaxy}. Moreover, we follow \citet{k19b} and the recommendations in \citet{boubert2019lessons}, and impose the following quality criteria on the \textit{Gaia} data considered in this analysis: (1) number of visibility periods $\geq 6$, (2) number of RV transits $\geq 4$, and (3) re-normalized unit weight error (RUWE) $\leq 1.6$.\footnote{The traditional square root of the reduced chi-square (unit weight error) has a strong dependence on color and magnitude. These dependencies are removed using a re-normalization process in \cite{LL:LL-124}. The re-normalized unit weight error (RUWE) provides a more robust indicator of the goodness-of-fit for the astrometry.} 

These quality cuts and the geometric selection described earlier results in a catalog of $\sim 1.7 \times 10^6$ stars. The median uncertainties in parallax and the proper motions ($\sigma_{\rm{\mu_{\alpha^{*}}}}, \sigma_{\rm{\mu_{\delta}}}$) are $0.037$ mas, $0.06$ mas yr$^{-1}$, and $0.05$ mas yr$^{-1}$ respectively. The velocity error budget is dominated by the radial velocity (RV) measurements; the median RV uncertainty in our selected subsample is $1.15$ km s$^{-1}$. 

\begin{figure*}
 \includegraphics[width=168mm]{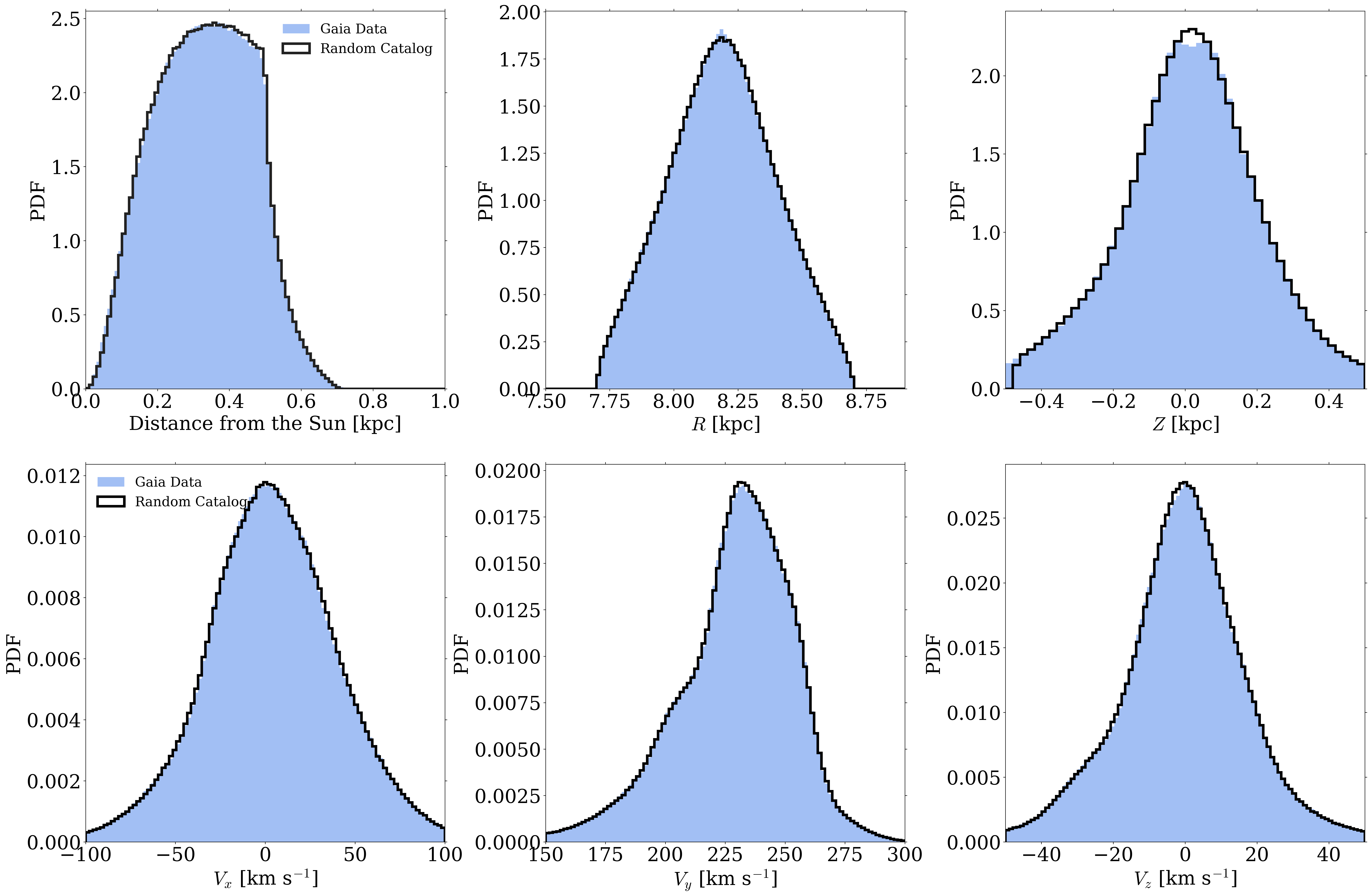}
 \caption{Validation of the method for random catalog construction by comparison to \textit{Gaia} data. In each panel the distribution of {\it Gaia} data (blue) is compared to the random catalog (black).  Top panels: Comparison in the quantities of heliocentric distance, and Galactocentric $R$ and $Z$ distributions. Bottom panels: Comparison in three components of the velocities: $V_x$, $V_y$, and $V_z$. Overall, the random catalog reproduces the distribution of the data in the solar neighborhood.}
 \label{fig:model}
\end{figure*}

\subsection{Simulations}
We use four simulations to interpret the \textit{Gaia} results, three from \cite{k19a} and one from \cite{rybizki2018gaia} (hereafter R18).\\
\subsubsection{\cite{k19a}}

K19a presented three simulations that we will utilize in this work. These simulations are summarized below; we refer the reader to K19a for a more comprehensive overview of the different model ingredients. 

\begin{itemize}

\item \textit{A fiducial simulation with both clustered star formation and a realistic gravitational potential.} The fiducial simulation self-consistently evolves 4 billion stars over the last 5 Gyr in a realistic time-varying potential that includes an axisymmetric component, a bar, spiral arms, and live giant molecular clouds (GMCs). All stars are initialized in clusters with an observationally-motivated range of initial conditions. For stars older than $5$ Gyr, we include a smooth, phase-mixed background population of stars. We developed a method of initializing star clusters to mimic the effects of direct \textit{N}-body interactions, while the actual orbit integrations are treated as test particles within the analytic potential.

\item \textit{A simulation with only small-scale perturbations.} The setup of this simulation is almost identical to the fiducial simulation with one key difference: the potential is axisymmetric with no bar and spiral arms, and no GMCs as pertubers.

\item \textit{A simulation with non-axisymmetric perturbations (with bar \& spiral arms) but with no clustered star formation (NCSF simulation hereafter).} Instead of forming stars within clusters, we form them as above but in $N=1$ systems. Since there are no clusters in this simulation, there is little small-scale clustering. 

\end{itemize}

To enable a fair comparison to the \textit{Gaia} data we create mock catalogs of our simulation in {\it Gaia} DR2-like solar cylinders. The solar neighborhood cylinder is centered at $(-8.2, 0.0, 0.025)$ kpc \citep{bland2016galaxy}, and has a radius of $0.5$ kpc and height of $1$ kpc (0.5 kpc above and below the solar position). We use the \texttt{MIST} stellar evolutionary tracks \citep{choi2016mesa} and the C3K stellar library (Conroy et al., unpublished) to derive photometry for the simulated stars using a Kroupa IMF \citep{kroupa2001variation}. We also calculate $G_\mathrm{RVS}$ using the relations (Equations 2 and 3) presented in \citet{brown2018gaia} and apply the same  $G_\mathrm{RVS}$ selection ($G_\mathrm{RVS} < 12$). 

An accurate error model is essential for comparisons between simulations and observations. The dependence of parallax, proper motion and radial velocity errors is a complex function of several parameters. We fit a Gaussian mixture model (GMM) with 20 components to the combined ($G, G_{\rm{BP}}-G_{\rm{RP}}, \sigma_{\varpi}, \sigma_{\rm{\mu_{\alpha^{*}}}}, \sigma_{\rm{\mu_{\delta}}}$) and ($G, G_{\rm{BP}}-G_{\rm{RP}}, \sigma_{\rm{RV}}$) spaces respectively, where $\sigma_{\varpi}, \sigma_{\rm{\mu_{\alpha^{*}}}}, \sigma_{\rm{\mu_{\delta}}}, \sigma_{\rm{RV}}$ are the uncertainties in the parallax, proper motions and the radial velocities. To draw realistic error estimates given $G$ and $G_{\rm{BP}}-G_{\rm{RP}}$ in our simulation, we sample from the conditional distributions for the respective errors given $G$ and $G_{\rm{BP}}-G_{\rm{RP}}$. However, the errors will also depend on the scanning law, which is not explicitly modelled for this work. The scanning law also has a non-trivial impact on the selection of stars in the solar neighborhood. With recent progress on modelling the scanning law and computing the true RVS selection function \citep{boubert2020completeness}, we plan to include a detailed selection function in future work (as opposed to a simple magnitude cut) for both the error model and the selection of stars. 

Since the simulations in K19 and R18 do not include binarity, it is important to avoid contamination from bound and disrupting wide binaries in the data. The Jacobi radius, beyond which the Galactic tidal field is stronger than the mutual gravitational attraction of binaries, is $\approx 1.7$ pc for $\sim 1 M_{\odot}$ stars in the solar neighborhood \citep{yoo2004end, jiang2010evolution}. We employ a simple condition on the projected separation between pairs of stars ($s < 0.5$ pc) to ensure minimal contamination from wide binaries. We discuss our motivations for this selection in Appendix B. Moreover, simulations in \citet{jiang2010evolution} also predict that there could be a noticeable signature of unbound wide binaries in the phase space density up to $10-100 R_J$ ($\sim 20-200$ pc), where $R_J$ is the Jacobi radius. A discussion on unbound wide binaries is also included in Appendix B.

\subsubsection{\cite{rybizki2018gaia}}

The smooth, unclustered \textit{Gaia} DR2 mock catalog presented in R18 is essential to validate the techniques presented in this paper. The mock catalog in R18 was generated using Galaxia \citep{sharma2011galaxia}, sampling stars according to the (spatially and kinematically smooth) Besancon Galactic model \citep{robin2003synthetic}. Moreover, R18 also include a realistic treatment of 3D dust extinction \citep[][and references therein]{bovy2016galactic}, and \textit{Gaia} DR2-like errors in the astrometry, photometry, and spectroscopy of stars. The R18 mock uses PARSEC isochrones \citep{marigo2017new} to generate the photometry. Lastly, R18 also include a model for the Galactic warp \citep{sharma2011galaxia}. 

To enable a fair comparison to the \textit{Gaia} data we create a DR2-like solar cylinder from the R18 mock data. The solar neighborhood cylinder is centered at $(-8.2, 0.0, 0.025)$ kpc, and has a radius of $0.5$ kpc and height of $1$ kpc (0.5 kpc above and below the solar position). To mimic the $G_\mathrm{RVS}$ selection, we calculate $G_\mathrm{RVS}$ using the relations provided in \cite{brown2018gaia} (Equations 2 and 3) and make the selection $G_\mathrm{RVS} < 12$. The R18 mock has no spatial or kinematic clustering on any scales by construction; consequently, we will use the R18 mock as a control to test our technique for measuring the TPCF, and validate our technique to generate random catalogs. 

\section{Two-Point Correlation Function}
\label{sec:tpcf}
\subsection{Theory and Motivation}

The two-point correlation function (TPCF) is a powerful measure of the clustering in data \citep[see][and references therein]{peebles2001galaxy}.  The TPCF measures the excess probability of finding one object within a specified distance of another object against that of a random, unclustered distribution. In the sections that follow, we will calculate the TPCF using a variety of different weighting schemes to implement cuts on $\Delta v$, $\Delta r$, $\Delta$[Fe/H], or $\Delta$age. We use the highly-optimized, OpenMP-parallelized, publicly-available code, \texttt{CorrFunc}\footnote{https://github.com/manodeep/CorrFunc} \citep{sinha2018corrfunc, sinha2020corrfunc}, to calculate the TPCFs. 

\begin{equation}
	dP_{1,2} = n^2 dV_1 dV_2 [1 + \xi(\Delta r_{1,2})].
\end{equation}

\noindent Here, $n$ is the mean density, and $\xi(\Delta r_{1,2})$ is the excess probability, relative to an unclustered distribution, that two points (1 and 2) are separated by $\Delta r_{1,2}$ ($\Delta r$ hereafter\footnote{In other fields, this is sometimes denoted as $r$ but we use $\Delta r$ to avoid confusion with the overall geometry of the system.}). The two-point correlation is usually estimated using the Landy-Szalay estimator \citep{landy1993bias}:

\begin{equation}
    \xi(\Delta r) = \frac{DD(\Delta r) - 2DR(\Delta r) + RR(\Delta r)}{RR(\Delta r)},
\end{equation}

\noindent where $DD$ is the count of data-data pairs, DR is the count of data-random pairs, and $RR$ is the count of random-random pairs. In studies of large-scale structure, a simple uniform distribution is adopted for the random catalog. However, the non-uniform density distribution of the Galactic disk makes the creation of the random catalog highly non-trivial \citep[e.g.,][]{mao2015probing}.

\begin{figure}
 \includegraphics[width=84mm]{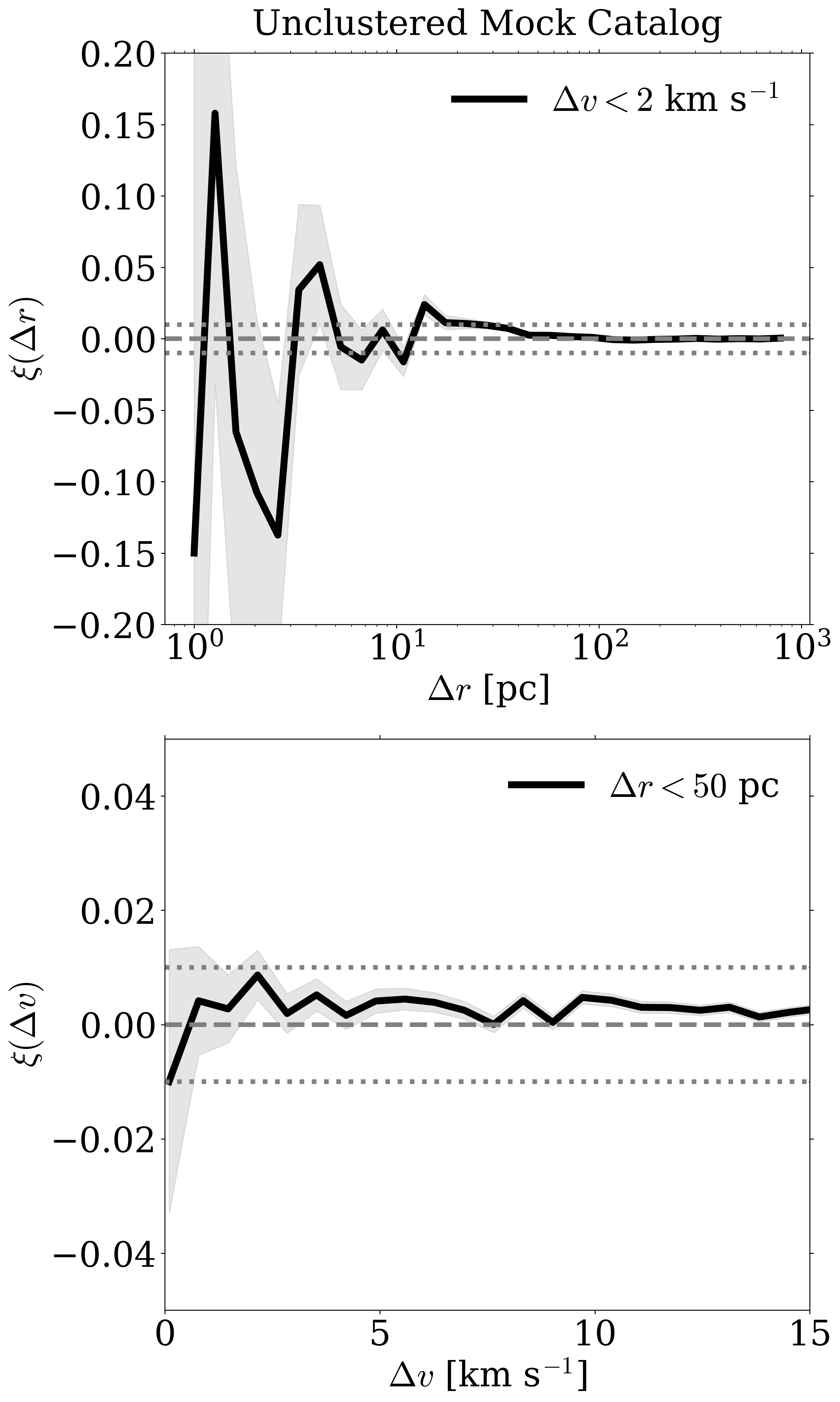}
 \caption{Correlation function measured in a mock dataset (R18) that was constructed to be smooth in position and velocity space.   \textit{Top panel:} the spatial TPCF for the R18 mock with $\Delta v< 2$ km s$^{-1}$. The shaded grey region shows the Poisson error.  \textit{Bottom panel:} the kinematic TPCF for the R18 mock with $\Delta r< 50$ pc. The shaded grey region shows the Poisson error. In both panels there is no statistically significant clustering above $10^{-2}$ at any scale. The absence of any signal in the TPCF for the R18 mock indicates the fidelity of the technique used to generate the random catalogs and measure the correlation function. }
 \label{fig:rc}
\end{figure}

\subsection{Random Catalogs}

Random catalogs are an essential ingredient for computing the TPCF. \cite{lancaster2019quantifying} utilized the TPCF to quantify the smoothness of the Galactic halo; the random catalog was created by fitting a parametric form of the density profile to the halo. \cite{mao2015probing} (M15 hereafter) used the TPCF to probe Galactic disk structure in SEGUE G-Dwarf stars. The results presented in M15 show that both the underlying density gradient and the survey geometry can significantly bias the TPCF. M15 chose analytical thin-disk/thick-disk density models, assuming that the $\xi(\Delta r)$ should approach 0 at sufficiently large scales if the correct density model is chosen. The key result from M15 was providing constraints on the scale-lengths and the scale-heights of both the thin disk and the thick disk. 

The combination of a non-uniform density profile and a complex selection function for the data necessitate the use of a very flexible density estimation technique. For this work, we attempted Gaussian mixture models, Dirichlet process Gaussian mixture models (also known as infinite mixture models; \cite{rasmussen2000infinite}), and normalizing flows \citep[e.g.,][]{rezende2015variational}. In an attempt to balance accuracy and interpretability, we chose to use Dirichlet process Gaussian mixture models (DPGMM hereafter). 

Gaussian mixture models have been extensively used in astronomy. GMMs assume that the input data are generated from a mixture of a finite number of Gaussian distributions with unknown parameters. Traditionally, GMMs are trained using the expectation-maximization algorithm and require choosing hyperparameters. The GMM can be written as: 

\begin{equation}
	p(x | \theta_1, ... , \theta_K) = \sum_{k=1}^K \pi_k \mathcal{N}(x | \mu_k, \Sigma_k) \\
\end{equation}

\begin{equation}
    \mathcal{N}(x | \mu, \Sigma) = \frac{1}{(2\pi)^{D/2}} \frac{1}{|\Sigma|^{1/2}} \exp \left(-\frac{1}{2}(x-\mu)^T \Sigma^{-1} (x-\mu)\right),
\end{equation} 

\noindent where $\theta_k = \{\mu_k, \Sigma_k, \pi_k\}$ is the set of parameters for component $k$, and $\mathcal{N}(x | \mu, \Sigma)$ is the multivariate Gaussian. 

The Dirichlet process Gaussian mixture model (DPGMM) is a non-parametric, Bayesian extension of GMMs, where each parameter in the model is assigned a prior. The DPGMM is formally written as:

\begin{eqnarray}
(\mu_k, \Sigma_k) &\sim& \mathcal{N}\mathcal{I}\mathcal{W}(\mu_0, \lambda_0, S_0, \nu_0)\\
\pi_k &\sim& \mathcal{D}\mathcal{P}(\alpha) .
\end{eqnarray}

\begin{figure}
 \includegraphics[width=84mm]{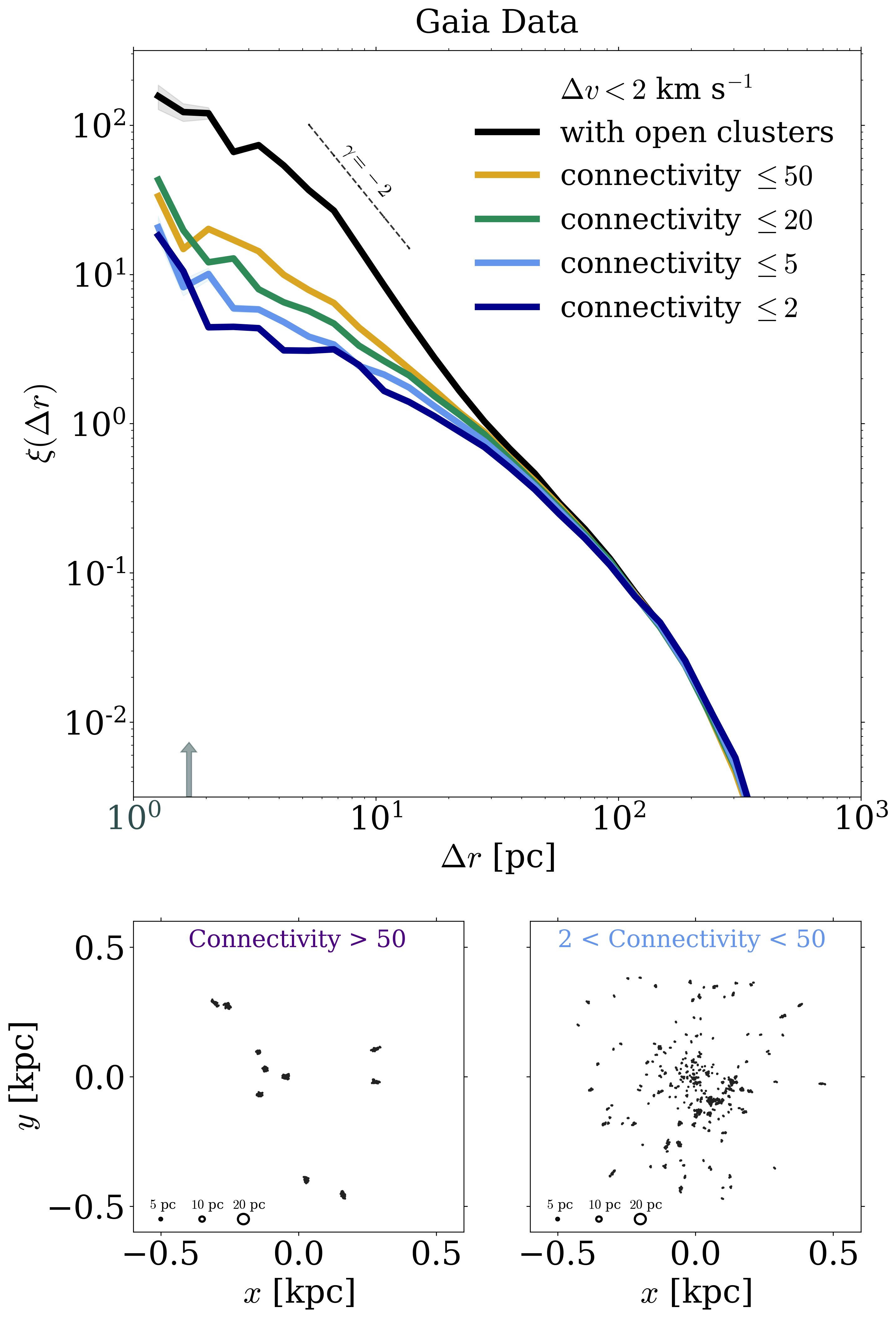}
 \caption{TPCF for the \textit{Gaia} data with $\Delta v < 2$ km s$^{-1}$ and various connectivity selections, and an illustration of what these different selections look like spatially. \textit{Top panel:} Different colors show the spatial TPCF with open clusters and with varying connectivity cuts ($\leq \{2,5,10,50\}$) to remove bound structures. The Jacobi radius for wide binaries in the solar neighborhood is shown as an arrow on the x-axis. Black line shows the power-law index $\gamma = -2$. The TPCF at scales $< 30$ pc for no connectivity cut is largely dominated by stars from a few bound open clusters. As the connectivity cut gets smaller, there is not a large difference between a connectivity of $\leq 20$ and $\leq 2$. For the rest of this work, we choose stars with a connectivity of $\leq 5$ for both the data and the simulations. \textit{Bottom left:} stars in the solar neighborhood with a connectivity of $> 50$. Most of the stars shown here are a part of a bound open cluster. \textit{Bottom right:} stars with a connectivity between $2$ and $50$. These stars are may be a part of a larger star-forming region, stars that are subsampled from small or more diffuse open clusters, or chance alignments.}
 \label{fig:conn}
\end{figure}

\begin{figure*}
\begin{center}
 \includegraphics[width=0.8\textwidth]{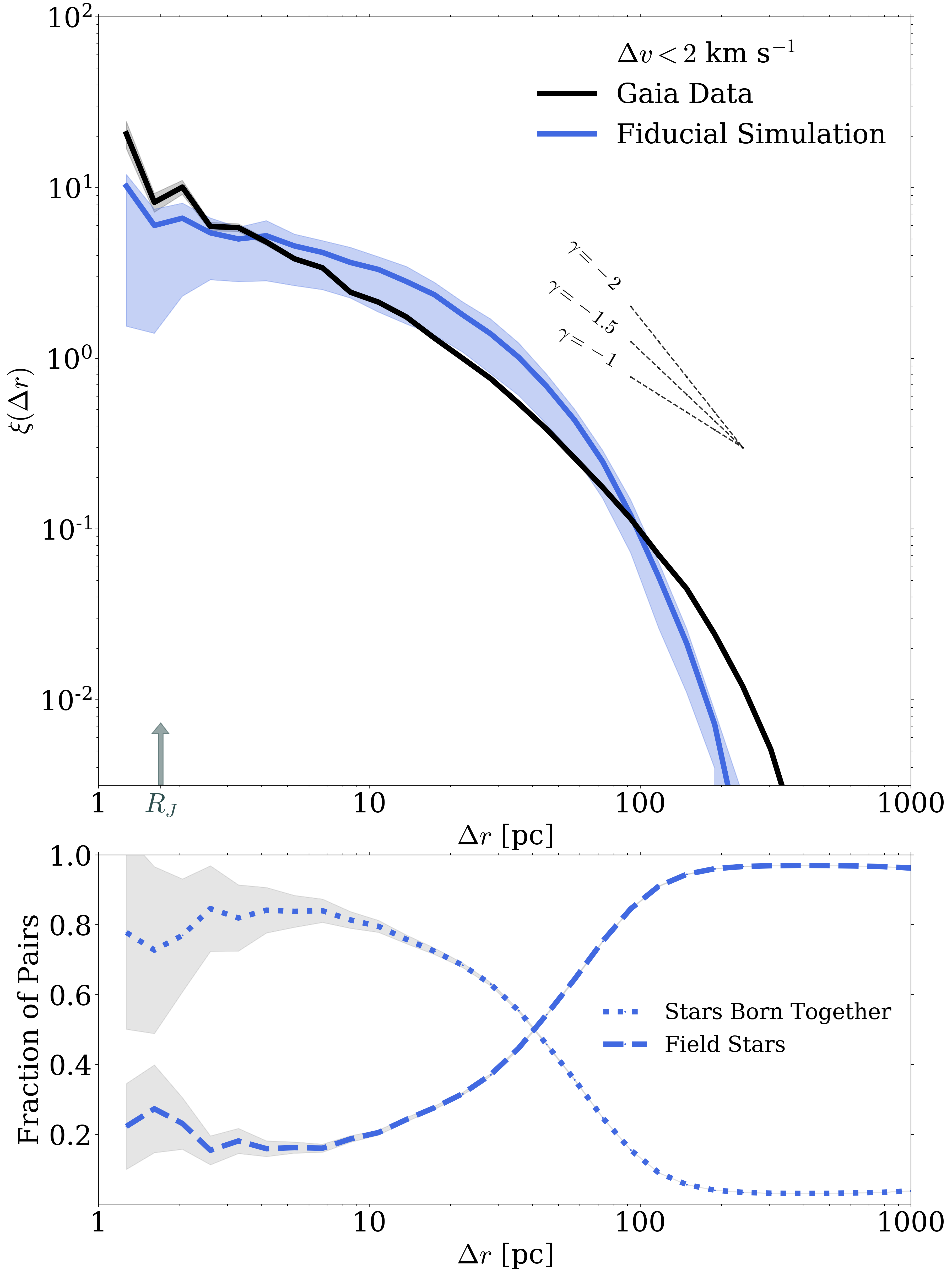}
 \caption{\textit{Top panel:} TPCF for the \textit{Gaia} solar cylinder (black) and the fiducial simulation (blue) with a 3D velocity difference of $2$ km s$^{-1}$. Poisson errors for the data are shaded in black. The blue shaded region shows the sample variance for the simulation (calculated by measuring the TPCF for different ``solar neighborhoods'') and the poisson are error added in quadrature. An arrow shows the Jacobi radius, $R_J$, of solar-mass wide binaries in the solar neighborhood.  \textit{Bottom panel:}  Fraction of pairs in the simulation that were born together (blue dotted) and the fraction that are chance field alignments (blue dashed). At low $\Delta r$ (and $\Delta v < 2$ km s$^{-1}$), the fraction of stars born together dominate the simulation TPCF presented in the upper panel. Beyond $\sim 40$ pc the majority of pairs arise from chance field alignments.}
 \label{fig:main}
 \end{center}
\end{figure*}

\noindent Here, the means and the covariances of each Gaussian component ($\mu_k, \Sigma_k$) have the Normal Inverse Wishart (NIW) distribution as their prior. $\mu_0, \lambda_0, S_0, \nu_0$ represent the mean, scale, scale matrix, and the degrees of freedom. The weights for each Gaussian component follow a Dirichlet process prior, parameterized by the concentration $\alpha$. The key advantages that DPGMMs offer over GMMs are two-fold, (1) the number of components actively used in the model are automatically inferred using variational inference, and (2) the priors help regularize the model. We use the implementation of DPGMM's presented in scikit-learn \citep{pedregosa2011scikit}. The model is fit to the entirety of \textit{Gaia} RVS because the rigid boundaries of a cylindrical selection created some pathological behavior in the fitting process. The model is fit on half the data and compared to the other half for validation. We first sample stars from the fit DPGMM, and then make the spatial selection described in Section 2. 

The fidelity of the random catalogs generated using the DPGMM method described above is shown in Figure \ref{fig:model}. The top left panel of Figure \ref{fig:model} shows the distribution of the distance of each star in the \textit{Gaia} solar cylinder and the random catalog to the solar position. The distance distribution is probing the spatial density distribution of stars in the solar cylinder subjected to the \textit{Gaia} selection function -- both the data and the random catalog appear to be in excellent agreement. The top, middle, and right panels of Figure \ref{fig:model} show the distribution of $R$ and $Z$ in the data and the random catalog. The bottom three panels show the distribution of the three velocity components for the data and the random catalog. The data and the random catalog seem to be in good agreement for the different spatial and velocity components. Consequently, the random catalogs created here generate an unclustered distribution of stars that adequately reproduce both the phase space distribution and the impact of the selection function of \textit{Gaia} stars. We use the same method to generate the methods for the R18 mock, the K19a simulations, and the \textit{Gaia} data. 

A much more rigorous test of the random catalog construction is to apply our machinery to the R18 mock catalog, which should have no clustering signal.  The resulting TPCF is shown in Figure \ref{fig:rc}.  The top panel shows $\xi (\Delta r; \Delta v < 2$ km s$^{-1})$ and the bottom panel shows $\xi(\Delta v; \Delta r < 50$ pc$)$.  There is no clustering signal at a level exceeding $10^{-2}$, indicating that our approach to measuring the TPCF is reliable at this level.  We note that the R18 mock contains both a Galactic warp and spatially inhomogenous dust. Clearly these two physical effects do not have any effect on the measured correlation function.

\subsection{Identifying and Removing Bound Structures}

Calculating the TPCF is an exercise in pair-counting. Large open clusters that are close-by and well-sampled in the data could dominate the TPCF because the pair counts scale as $N^2$. There has been a significant amount of effort toward finding and characterizing these large open clusters \citep[e.g.,][]{de2009hierarchical}, and many new open clusters are being found with \textit{Gaia} DR2 \citep[e.g.,][]{cantat2018gaia, castro2020hunting}. It is easy to see that known open clusters could overwhelm the TPCF signal at small spatial scales. 

Consequently, we choose to exclude pairs from these open clusters to isolate the signal of star clusters that are disrupting or have already disrupted. Similar to \citet{oh2017comoving} and \cite{k19b}, we form an undirected graph where stars are nodes, and edges between the nodes exist for co-moving pairs of stars. For the purposes of this work, we define co-moving as having a 3D velocity difference of $<2$ km s$^{-1}$ and a physical separation of $<5$ pc; these selections are similar to linking lengths in the Friends-of-Friends algorithm. Consequently, a star could have multiple co-moving neighbors, and a pair of stars could be directly or indirectly connected via a sequence of edges. The graph is then split into connected components -- a connected component is a subgraph of the original graph in which any two nodes are connected to each other by a path -- to calculate the connectivity of each star. A connectivity of 1 means that a star is not a part of any larger structure, a connectivity of 2 means that a star is in a mutually exclusive pair, and so on. 

Bound wide binaries could also impact the TPCF at small spatial scales. The actual separations for most of these bound wide binaries is likely less than the Jacobi radius in the solar neighborhood ($\sim 1.7$ pc). However, the median parallax uncertainty of $\sim 0.04$ mas corresponds to an uncertainty of $\pm 10$ pc at 500 pc. Consequently, to minimize contamination from bound wide binaries, we impose the additional condition that all TPCF calculations for the data exclude pairs that have projected separation $< 0.5$ pc. A discussion of why we choose this criterion is included in Appendix B. 

The top panel of Figure \ref{fig:conn} shows the spatial TPCF calculated  for stars with $\Delta v < 2$ km s$^{-1}$ and with different connectivity cuts in the \textit{Gaia} data.  The bottom-left panel shows all the stars with a connectivity of $>50$. This selection efficiently identifies nearby open clusters; the clusters picked out above include Melotte 20, Pleiades, NGC 2516, Hyades, and Praesepe. The right panel shows stars with a connectivity between $2$ and $50$. 

The TPCF that includes all stars (including stars from open clusters) shows the strongest clustering in the data. Even with a liberal connectivity cut where stars that are part of connected components with a size of 20 or less are included, the TPCF is notably stable and close to what it is with the very conservative connectivity $\leq 2$ cut, which only selects unconnected (connectivity $=1$) and mutually exlusive stars (connectivity $=2$). The TPCF being largely insensitive to smaller connectivity cuts is reassuring as it indicates that the bound open clusters are being effectively filtered out. Consequently, we choose to use a connectivity $\leq 5$ selection for the rest of this work. 

\section{Results}
\label{sec:results}

\subsection{Comparing Data \& Simulations}
\label{sec:resultsp1}

In this section we present the spatial and kinematic TPCF in the data and compare to three simulations from K19a.

The top panel of Figure \ref{fig:main} shows the TPCF in the data and the fiducial simulation for stars with a 3D velocity difference of less than $2$ km s$^{-1}$. The Jacobi radius ($R_J$) of wide binaries in the solar neighborhood is shown as an arrow on the x-axis; there could be contamination from bound wide binaries at separations smaller than $R_J$. Poisson uncertainties are shown as shaded bands. The three dashed lines show different curves with power-law index $\xi(\Delta r) \propto \Delta r^{\gamma}$, $\gamma=-1.0,-1.5,-2.0$.

\begin{figure*}
 \includegraphics[width=168mm]{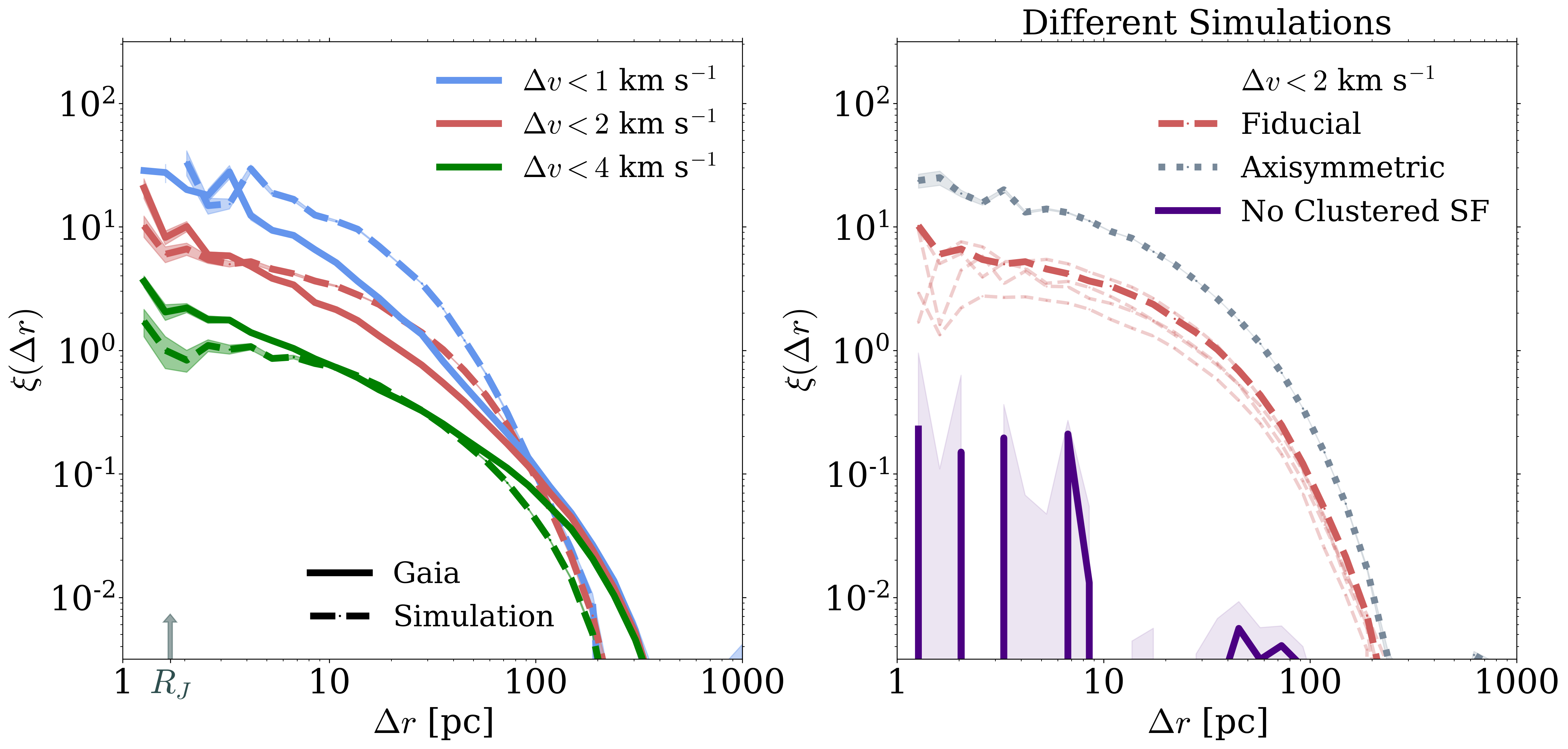}
 \caption{\textit{Left panel: } TPCF with $\Delta v = \{1,2,4\}$ km s$^{-1}$  (blue, red, green) for the data (solid lines) and the simulation (dashed lines). Each curve has associated Poisson errors shaded. The features at different spatial scales discussed in Figure 5 for $\Delta v < 2$ km s$^{-1}$ are also apparent at different velocity amplitudes. The data seem to diverge the most from the simulation at $\Delta v < 1$ km s$^{-1}$ at intermediate scales. The discrepancies between the data and the fiducial simulation for different $\Delta v$ indicates the need for a more complex subgrid model of star formation and the non-axisymmetries in the potential. \textit{Right panel:} TPCF for the fiducial (red), axisymmetric (gray), and the NCSF simulation (purple). The light red lines near the fiducial TPCF show the TPCF in four different solar cylinders throughout the simulated galaxy. As expected, the axisymmetric simulation is more clustered due to no scattering and the NCSF simulation is less clustered due to the absence of clustered star formation. The sample variance in the different solar neighborhoods contributes a factor of $2-3$ variation in the TPCF.}
 \label{fig:diff_dv}
\end{figure*}

There are three spatial regimes to consider: small ($<10$ pc), intermediate ($10 < \Delta r < 100$ pc), and large ($100 < \Delta r < 1000$ pc). The data are slightly more clustered than the fiducial simulation at the smallest scales by a factor of $\sim 2-3$. The fiducial simulation initializes all stars into star clusters \citep{lada2003embedded}, and explicitly undervirializes the stars to mimic the boundedness of stars born together at young ages. The data could include some stars from subsampled bound open clusters at these small spatial scales that would not have an analog in the simulation. 

At intermediate scales, the simulation and the data are in reasonable agreement (see Figure 6 for a discussion on sample variance). Clustering at these scales is likely caused by disrupting star clusters that are still overdense in phase space or unbound stellar associations \citep[e.g.,][]{meingast2019extended}. The addition of a more diffuse (or hierarchical) mode of star formation \citep{kruijssen2012fraction} could help explain some of the discrepancy at these intermediate and large scales. 

The data are again more clustered than the simulation at the largest scales. Clustering at these scales could either be caused by the few pairs of stars that were born together but drifted apart or by the resonances related to the non-axisymmetries of the Galactic potential. A hierarchical model for star formation could explain the clustering at large $\Delta r$ since we would expect different star-forming regions to also be spatially correlated \citep{grasha2017hierarchical}. The fiducial simulation includes a realistic bar and rotating but fixed spiral arms; consequently, the disagreement at larger $\Delta r$ could also indicate clustering due to resonances the transient spiral arms. As shown in previous work \citep[e.g.,][]{hunt2018transient, sellwood2019discriminating}, the inclusion of transient spiral arms will likely show richer phase space structure in the solar neighborhood at large scales.

\begin{figure}
 \includegraphics[width=84mm]{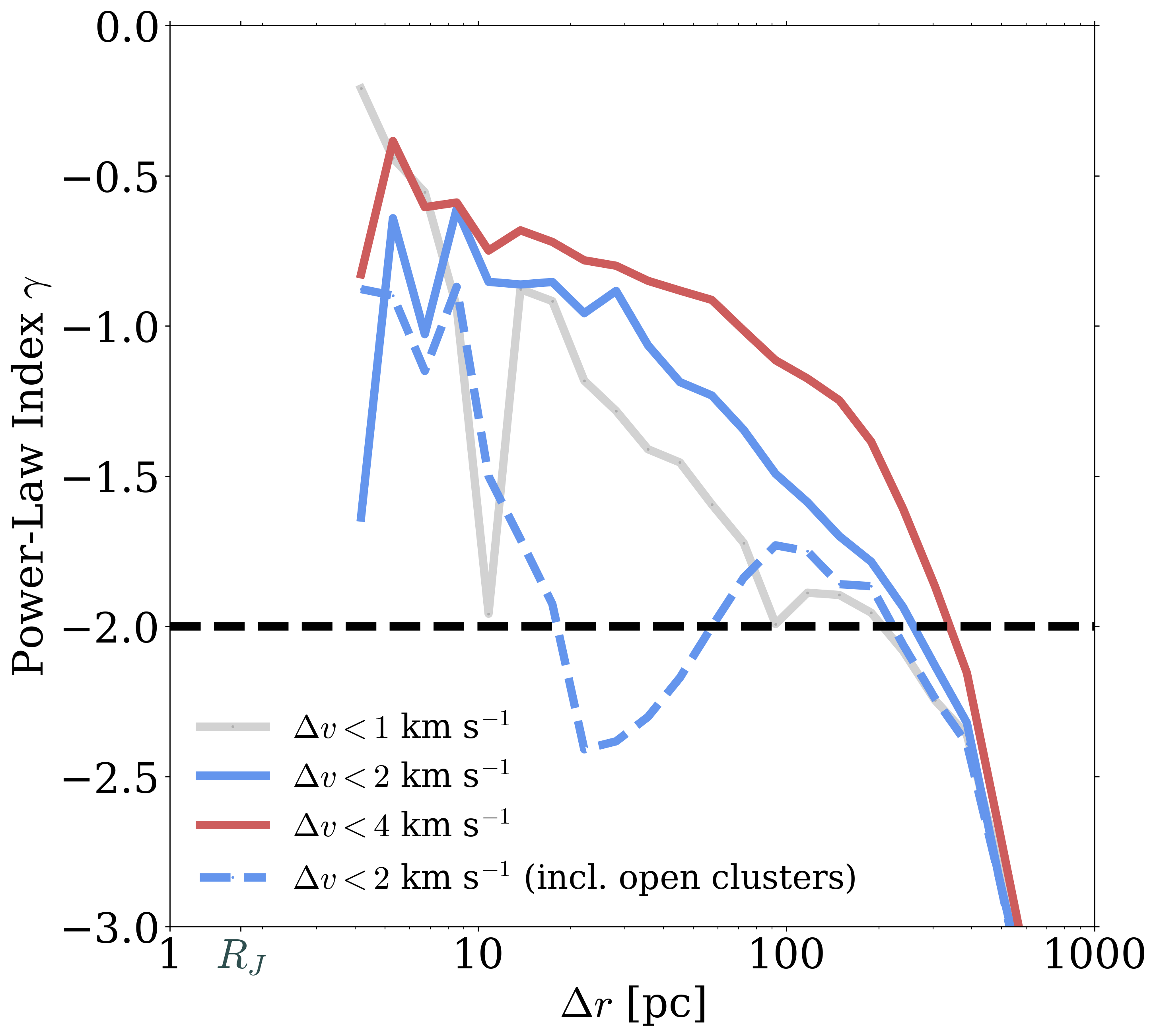}
 \caption{Power-law index ($\xi(\Delta r) \propto \Delta r^{\gamma}$) for the \textit{Gaia} spatial TPCF with different $\Delta v$ cuts. We compute the average power-law index on a rolling basis. For $\lesssim 50$ pc, the index is $\sim -1$ for all $\Delta v$. Regardless of $\Delta v$, the index falls precipitously beyond $50-70$ pc to $\lesssim -2$. The power-law index for the data that includes bound open clusters (no connectivity cut) is $\sim -2$ for $\Delta r > 10$ pc. }
 \label{fig:pl}
\end{figure}

\begin{figure*}
 \includegraphics[width=168mm]{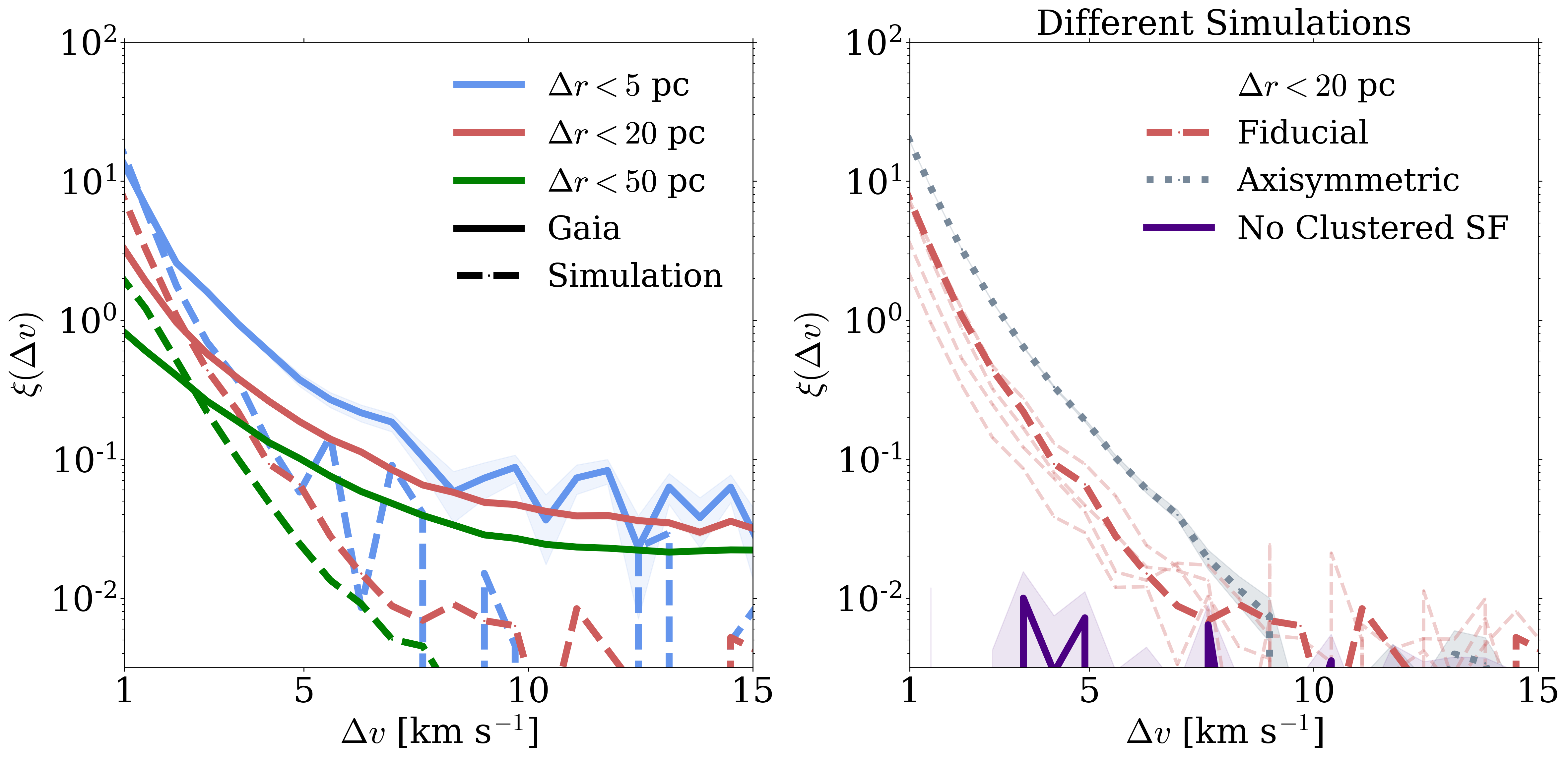}
 \caption{{\it Left panel}: velocity TPCF with $\Delta r = \{5,20,50\}$ pc (blue, red, green) for the data (solid lines) and the simulation (dashed lines). Each curve has associated Poisson errors shaded. The simulation and the data seem to be reasonably close at low $\Delta v$ (up to $\sim 3-4$ km s$^{-1}$); however, for all larger $\Delta v$, the data are more clustered than the simulation by a few factors regardless of the spatial scale. This rich velocity structure regardless of spatial scale likely points to both a more complex cluster dispersion initial velocity profile and the impact of resonances in the disk. {\it Right panel}: velocity TPCF for the fiducial (red), axisymmetric (gray), and the NCSF simulation (purple). The light red lines near the fiducial TPCF show the TPCF in different solar cylinders throughout the simulated galaxy. The sample variance for the kinematic TPCF is comparable to that of the spatial TPCF shown in Figure \ref{fig:main}. }
 \label{fig:diff_ds}
\end{figure*}

The bottom panel of Figure \ref{fig:main} shows the fraction of pairs in the simulation born together and the fraction of field pairs as a function of their spatial separation for a velocity difference of $\Delta v < 2$ km s$^{-1}$.  The fraction of stars born together dominates the pair counts up to $\sim 40$ pc. Field pairs dominate above for larger spatial scales. Consequently, the small spatial scale ($<10$ pc) probe clustered star formation, the intermediate scales ($10 < r < 100$ pc) probe clustering due to both star formation and other clustering mechanisms, and the large scales ($100 < r < 1000$ pc) mostly probe non-SF related clustering in the simulation. 

Figure \ref{fig:main} simultaneously probes star formation at small scales, the disruption mechanism of star clusters, and the resonances in the Galaxy. However, the chosen velocity difference is only probing clustering for one $\Delta v$ cut. Figure \ref{fig:diff_dv} shows the TPCF for the data and the three simulations from K19a for the velocity differences: $\Delta v < \{1,2,$ and $4\}$ km s$^{-1}$. The left panel shows the data (solid line) and the fiducial simulation (dashed line) for the three different velocity differences. The shaded regions accompanying each curve show the Poisson errors. The pattern observed in Figure \ref{fig:diff_dv} for $\Delta v<2$ km s$^{-1}$ also holds for $\Delta v<1$ and $4$ km s$^{-1}$. 

The data and simulation largely agree at small scales, the simulation consistently shows more clustering at intermediate scales, and the data are more clustered at the largest scales. The discrepancy between the data and the simulation at small spatial scales for $\Delta v < 4$ km s$^{-1}$ could suggest that the birth velocity dispersion for stars born together in the data is broader than than the model used in the simulation. The birth velocity dispersion for star clusters in the simulation is determined by assuming some potential for the star cluster and drawing from the cluster birth mass-radius relation. The latter is particularly uncertain \citep[e.g.,][]{parmentier2011puzzle}, but recent work \citep{choksi2019initial} has made progress in estimating a physically-motivated mass-radius relation. Similar to $\Delta v < 2$ km s$^{-1}$, the difference at intermediate and large scales could be attributed to a hierarchical mode of star formation, resonances due to transient spiral arms, or a combination thereof. 

The right panel shows the TPCF for the three different simulations (fiducial, red; axisymmetric, grey; and NCSF, indigo) for $\Delta v < 2$ km s$^{-1}$. The NCSF simulation, as expected, has the lowest clustering amplitude because of the absence of clustered star formation. The NCSF simulation does have non-axisymmetries but this is not detectable in the TPCF at $\Delta v < 2$ km s$^{-1}$. The axisymmetric simulation has the largest clustering amplitude because of the absence of scattering from the bar, spiral arms, and GMCs, and hence star clusters disrupt the slowest. The different red lines near the fiducial simulation show the TPCF for different realizations of the ``solar" cylinder from the solar annulus in an effort to study sample variance. There is considerable scatter in the TPCF for the different solar cylinders -- the numbers vary by almost  a factor of $3-4$ at smaller scales. Such a large sample variance necessitates caution in comparing the data and the simulation, especially when the two are very similar. 

In Figure \ref{fig:pl} we calculate the power-law slope $\gamma$ for the different $\Delta v$ cuts by fitting the relation $\xi(\Delta r) \propto \Delta r^{\gamma}$ to the \textit{Gaia} data after making the connectivity selection discussed in Section 3.3 (solid lines) and without a connectivity cut for $\Delta v < 2$ km s$^{-1}$ (dashed). The slope is calculated for 5 bins on a rolling basis through the different spatial scales. For $\lesssim 50$ pc, the index hovers around $\sim -1$ for all $\Delta v$. Regardless of $\Delta v$, the power-law slope falls precipitously after $50-70$ pc to $\lesssim -2$. With no connectivity cut, the data is well-described by $\gamma \sim -2$ from $1-100$ pc. \citet{guszejnov2018universal} developed a model for scale-free fragmentation and showed that scale-free structure formation would generically lead to a correlation of the form $\xi (\Delta r) \propto r^{-2}$, which is what we observe in the data when we include open clusters. 

So far, we have focused on the spatial TPCF by selecting samples in a narrow $\Delta v$ range; we can also consider the converse and measure the velocity-space TPCF in a narrow $\Delta r$ range.  Figure \ref{fig:diff_ds} shows such measurements comparing the data and the fiducial simulation. There is clearly more clustering in data regardless of the spatial selection compared to the simulation. For $\Delta r < 5$ pc both the data and simulation are clustered at low $\Delta v$ (up to $\sim 5$ km s$^{-1}$); however, the data are more clustered at larger $\Delta v$ (up to $10$ km s$^{-1}$). If the signal at low $\Delta v$ is due to co-natal stars, the larger clustering in the data could indicate that the birth velocity dispersion of stars born together is larger in data than in the simulation. There is a similar trend, though less drastic, for $\Delta r < 20, 50$ pc. The data and the model are reasonably close at low $\Delta v$ but seem to diverge at large $\Delta v$. The co-natal fraction in the fiducial simulation at such large $\Delta v$ is quite low, even for $\Delta r < 5$ pc \citep{k19a}. Consequently, the discrepancy could be driven by the rich structure created due to either the non-axisymmetries of the Galaxy and/or non-equilibria phenomena in the disk \citep[e.g.,][]{laporte2018response}. A thorough test would involve calculating the TPCF for simulations with satellites (either isolated or cosmological), such as \cite{sanderson2018synthetic, laporte2018response}, and compare the velocity-space TPCF at large $\Delta v$. 

The right panel shows the velocity TPCF of the three simulation variants for $\Delta r < 20$ pc. The simulation with no clustered star formation (NCSF) has the lowest clustering amplitude, as expected. The NCSF simulation does have non-axisymmetries but this is evidently not detectable in the kinematic TPCF for $\Delta r < 20$ pc at any velocity scale. The axisymmetric simulation has the largest clustering amplitude because of the absence of scattering from the bar, spiral arms, and GMCs to disrupt the coherent motion of stars born together. The thin red lines show the TPCF for different realizations of the ``solar cylinder"  in the fiducial simulation.  As in Figure \ref{fig:main}, the sample variance is significant. 

We speculate that the discrepancies between the data and the simulations at large $\Delta v$ scales -- where the clustering should be largely driven by field pairs -- could be driven by either transient spiral arms or interactions with a satellite. Both simulations include a bar and steady-state spiral arms but do not include transient modes and interactions with external perturbers. Recent work \citep[e.g.,][]{hunt2018transient, sellwood2019discriminating} has shown that transient spiral arms can recreate velocity structure very close to what is observed in the data. Modelling external perturbers is more complex because of the need to run expensive \textit{N}-body simulations. 

The spatial and kinematic TPCF presented above reaffirm the need for a deeper look at the cluster disruption model in K19a, and a thorough treatment of the transient non-axisymmetries of the Galactic potential and a hierarchical model of star formation to study clustering at larger $\Delta r$ and $\Delta v$. 

\begin{figure}
 \includegraphics[width=84mm]{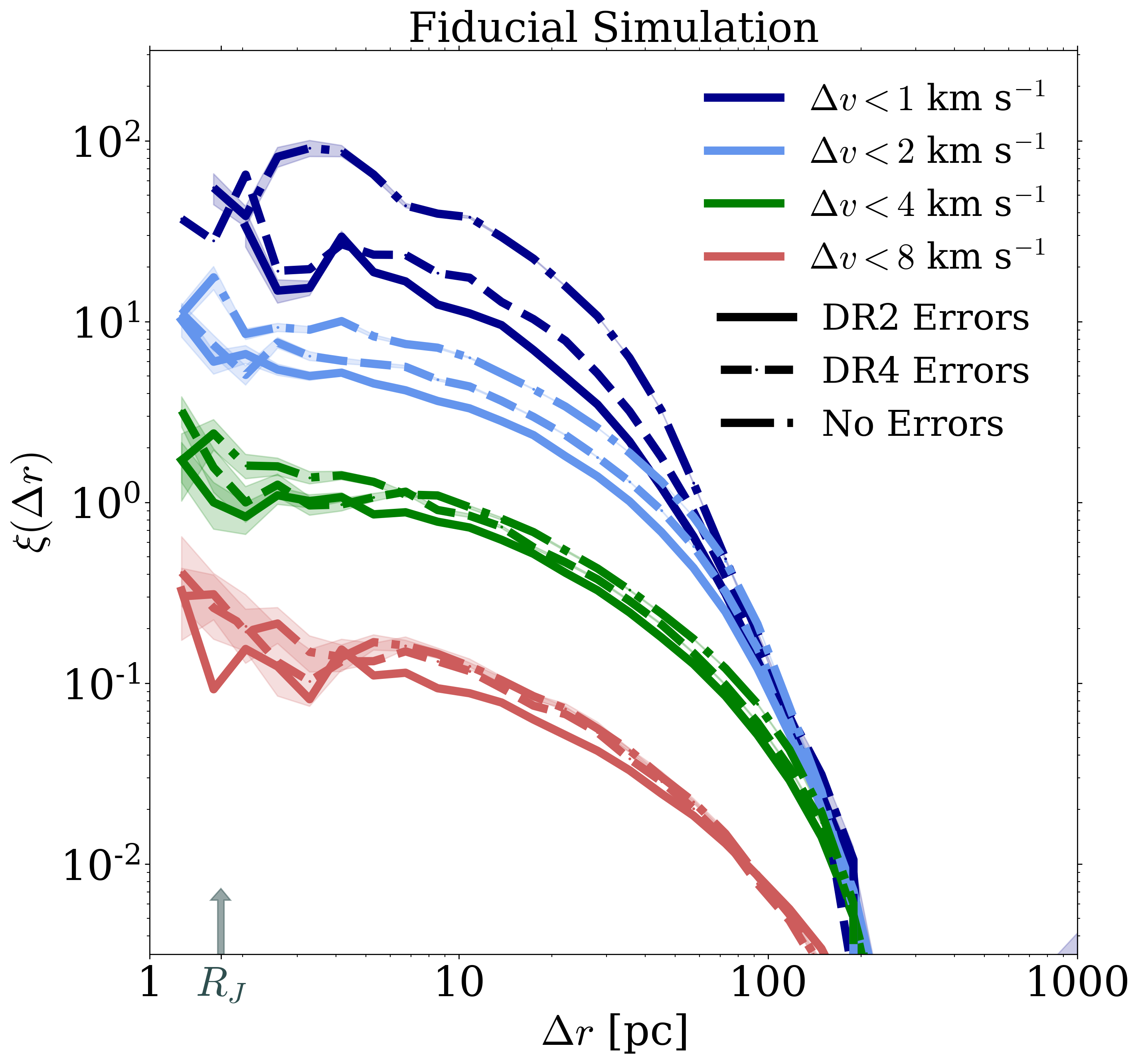}
 \caption{Impact of \textit{Gaia} DR2 errors, DR4 errors, and no phase space errors on the simulated $\xi(\Delta r; \Delta v < \{1,2,4,8\}$ km s$^{-1})$.  The improvement in precision leads to a modest change in $\xi(\Delta r)$ for $\Delta v < 4, 8$ km s$^{-1}$, and a notable increase in $\xi(\Delta r)$ for $\Delta v < 1,2$ km s$^{-1}$. }
 \label{fig:dp}
\end{figure} 

\begin{figure*}
 \includegraphics[width=168mm]{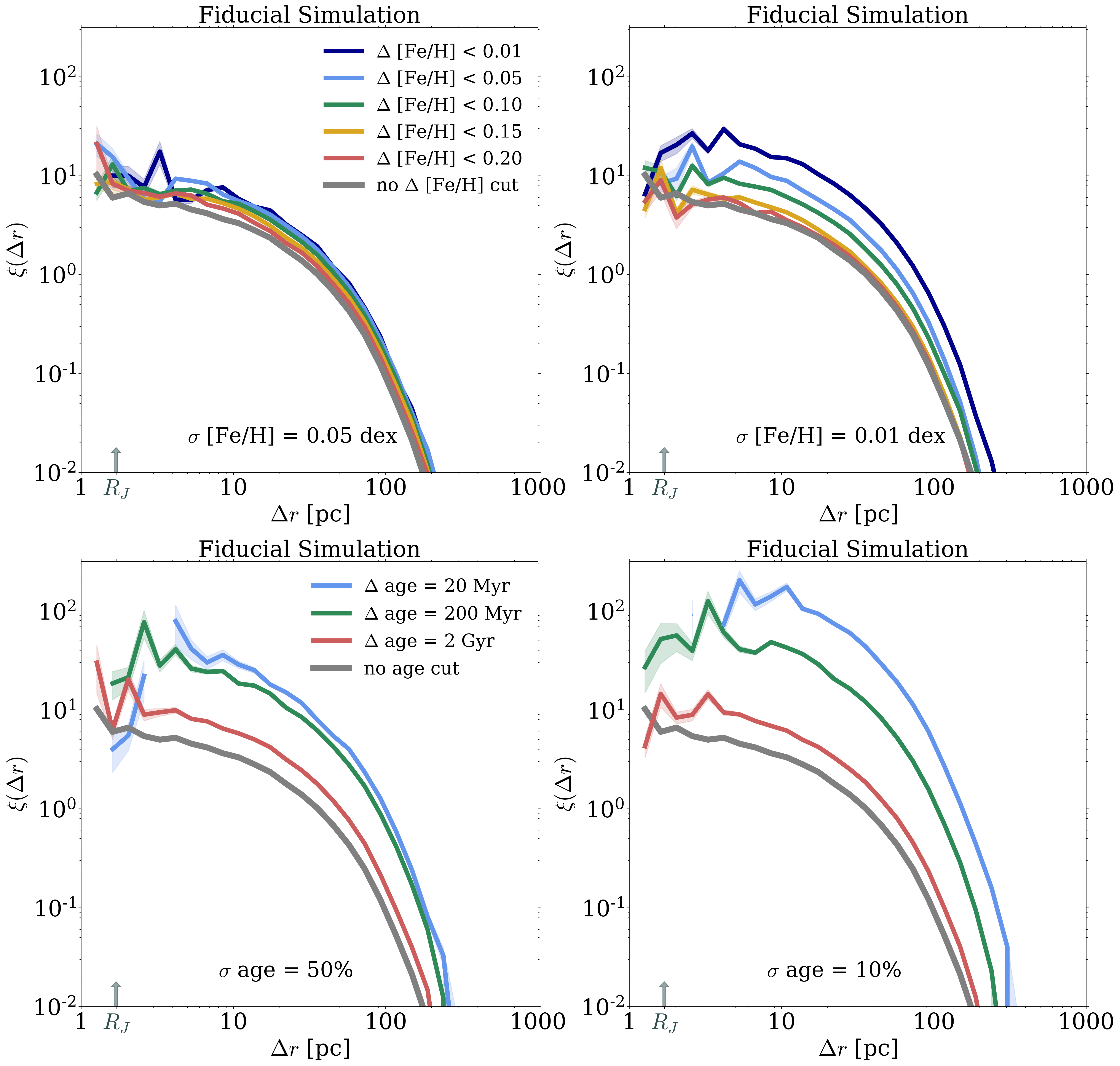}
 \caption{Change in $\xi(\Delta r; \Delta v < 2$ km s$^{-1})$ when including metallicities with $\sigma_{\rm [Fe/H]} = 0.01,0.05$ dex, and ages with $\sigma_\mathrm{age} = 10\%, 50\%$. {\it Top left}: $\xi(\Delta r; \Delta v < 2$ km s$^{-1})$ at various metallicity cuts with with $\sigma_{\rm [Fe/H]} = 0.05$ dex. The relative change in $\xi(\Delta r; \Delta v < 2$ km s$^{-1})$ with the addition of metallicity is about a factor of $\sim 1.7 - 2$. {\it Top right}: $\xi(\Delta r; \Delta v < 2$ km s$^{-1})$ at various metallicity cuts with with $\sigma_{\rm [Fe/H]} = 0.01$ dex (almost perfect metallicity). $\xi(\Delta r; \Delta v < 2$ km s$^{-1})$ with almost perfect metallicity information is an order of magnitude larger for the lowest $\Delta$ [Fe/H]. {\it Bottom left}: $\xi(\Delta r; \Delta v < 2$ km s$^{-1})$ with different $\Delta$age cuts and an age uncertainty of $50\%$. Even with an uncertainty of a factor of 2, the addition of ages leads leads to a similar change in $\xi(\Delta r; \Delta v < 2$ km s$^{-1})$ as the inclusion of almost perfect metallicity information. {\it Bottom right}: $\xi(\Delta r; \Delta v < 2$ km s$^{-1})$ with different $\Delta$age cuts and an age uncertainty of $10\%$. The change in $\xi(\Delta r; \Delta v < 2$ km s$^{-1})$ for the lowest $\Delta$ age cut is more than an order of magnitude. As expected, precise ages are very effective at identifying co-natal stars.}
 \label{fig:dfeh_dage}
\end{figure*}

\subsection{Predicted Clustering as a Function of {\it Gaia} Uncertainties, Age and Metallicity}
\label{sec:future}

The analysis so far has focused on the data products from \textit{Gaia} DR2, released in April 2018. The next decade will see an unprecedented increase in both the volume and precision of observational data of stars in the Galaxy. \textit{Gaia}, in particular, will deliver the radial velocities of tens of millions of stars, and will decrease the astrometric  uncertainties by at least a factor of 1.5\footnote{The \textit{Gaia} data improve with time as $t^{-0.5}$ for parallaxes, photometry, and radial velocities, and as $t^{-1.5}$ for proper motions (http://www.astro.lu.se/gaia2017/slides/Brown.pdf).}. The increase in precision and complementary spectroscopic and asteroseismic data prompt the question of how much the information content changes as new and more precise data become available. 

 We begin by exploring the effect of {\it Gaia} measurement uncertainties on the TPCF.  Figure \ref{fig:dp} shows the spatial TPCF for the fiducial simulation for DR2 errors (solid), DR4 errors (dashed), and for no errors (dash-dotted). To calculate the DR4 errors, we scale the parallax and the RV errors by $1/1.7$ and the proper motion errors by $1/4.5$. The DR4 $\xi(\Delta r)$ is not too different for $\Delta v<4,8$ km s$^{-1}$, which is not surprising given that the median DR2 uncertainties in our sample is 1.15 km s$^{-1}$. The $\xi(\Delta r)$ is expected to be approximately a factor of 1.5 times larger than the DR2 $\xi(\Delta r)$ for small $\Delta v$. The dash-dotted line shows the maximal change in $\xi(\Delta r)$ with perfect phase space information. As with DR4 errors, there is little difference at large $\Delta v$. However, the gain in $\xi(\Delta r)$ that can be extracted with perfect phase space information for low $\Delta v$ is more than a factor of 5 compared to \textit{Gaia} DR2 currently provides.

Alongside \textit{Gaia}, many ongoing \citep[e.g.,][]{deng2012lamost, de2015galah, kunder2017radial, 2017AJ....154...94M} and upcoming spectroscopic surveys \citep[e.g.,][]{de20124most, dalton2014project, kollmeier2017sdss} will deliver precise chemical information for millions of stars. The key challenge with modeling the multi-dimensional chemistry that these surveys will measure is grappling with the inherent dimensionality of the chemical space \citep[e.g.,][]{ting2012principal, price2018dimensionality}. Moreover, the uncertainties of the derived abundances are difficult to accurately forecast; as shown in \citet{ting2015apogee}, due to the large dimensionality of the chemical space, if the covariances are not taken into account, the effective uncertainty could be much smaller. Moreover, recent work \citep[e.g.,][]{martig2016red, ting2019vertical, bovy2019life} has shown the promise of utilizing age information to make inferences about dynamics in the Galactic disk. 

Figure \ref{fig:dfeh_dage} shows the potential impact of incorporating metallicity and age information when measuring the TPCF. The top panels show $\xi(\Delta r; \Delta v < 2)$ with various metallicity cuts for $\sigma_{\rm [Fe/H]} = 0.01,0.05$ dex. For $\sigma_{\rm [Fe/H]} = 0.05$ dex, the relative gain in information for the smallest $\Delta$[Fe/H] is a modest factor of $\sim 1.6-1.8$. The results for the TPCF line up well with those presented for the co-natal fraction of stars with additional metallicity information presented in K19a. The right panel shows the same $\xi(\Delta r; \Delta v < 2$ km s$^{-1})$ with various metallicity cuts but for an almost perfect [Fe/H] measurement with $\sigma_{\rm [Fe/H]} = 0.01$ dex. The change in $\xi(\Delta r)$ is almost an order of magnitude because the stringent $\Delta$ [Fe/H] selection is very efficient at identifying co-natal stars. 

The bottom panels of Figure \ref{fig:dfeh_dage} show $\xi(\Delta r; \Delta v < 2)$ with various different age cuts ($\Delta$ age = $20,200,2000$ Myr) for age uncertainties $\sigma_{\mathrm{age}} = 10\%$, $50\%$. Even with an age uncertainty of $50\%$, $\xi(\Delta r; \Delta v < 2)$ increases by an order of magnitude for the smallest $\Delta$ age cut of $20$ Myr. There are a few possible reasons for this large change with such uncertain ages. First, there is a much larger dynamic range in stellar ages compared to metallicities. Second, the uncertainties are relative, and so younger stars have smaller absolute uncertainties. Lastly, there is a stronger coupling between age and dynamics compared to the metallicity and dynamics; for instance, the age-velocity dispersion relation has a smaller scatter than the metallicity-velocity dispersion relation. Consequently, a weak prior on age and a strong prior on metallicity lead to an analogous change in the TPCF. The more precise ages ($\sigma_{\mathrm{age}} = 10\%$) increases $\xi(\Delta r; \Delta v < 2)$ by almost two orders of magnitude; precise ages combined with kinematics hold the most information about stars born together in the disk. 

\section{Summary}
\label{sec:conclusions}

Several key physical processes including the clustered nature of star formation, non-axisymmetries of the Galactic potential, and non-equilibria phenomena determine the structure of the Galaxy in chemodynamical space.
The two-point correlation function, a clustering metric widely used in other fields of astronomy and physics, is well-suited to the task of disentangling structure in chemodynamical caused by these distinct physical processes. 

In this paper presented a robust, non-parametric technique to generate realistic random catalogs for a complex density profile and a non-trivial selection function using Dirichlet Process Gaussian Mixture models. We validated the fidelity of our random catalog by calculating the TPCF of the R18 mock \citep{rybizki2018gaia}. We calculated the spatial and kinematic TPCF in the data and three simulations from \citet{k19a} sliced in velocity, separation, metallicity, and age. The resulting structure in these different contexts holds valuable clues about the nature of star formation and the importance of non-axisymmetries in the Galaxy. 

Our key findings are listed below. 

\begin{itemize}

    \item We calculate the spatial TPCF for stars with velocity ($\Delta v$) differences of $1,2,4$ km s$^{-1}$ and the kinematic TPCF for stars with spatial ($\Delta r$) separations of $5, 20, 50$ pc in the solar neighborhood with data from \textit{Gaia} DR2. We detect clustering out to large spatial and kinematic scale (up to $300$ pc and $15$ km s$^{-1}$). The power-law index of the spatial TPCF that includes bound structures is $\sim -2$ for $\Delta r > 10$ pc (in line with theoretical predictions), and is $\sim -1$ up to $50$ pc and then drops precipitously to $\lesssim -2$ for larger $\Delta r$ without bound structures.
    
    \item We analyze a novel star-by-star simulation \citep{k19a} to interpret the observational results. The data and the simulation agree reasonably well at small spatial scales but there is some tension at intermediate and large spatial scales. Since we assume in K19a that all stars are born in clusters (naturally leading to a more clustered population of stars), we suggest that the mismatch at intermediate scales could be explained by hierarchical star formation. For $\Delta v > 5$ km s$^{-1}$ and $> 100$ pc, the data show rich clustering in the spatial and kinematic TPCF that is absent in the simulations. Since the co-natal fraction in our simulations is small at these scales, we speculate that the low clustering strength in the simulations is due to the lack of transient spiral arms in the simulation. 

    \item Ongoing \textit{Gaia} data collection and upcoming spectroscopic surveys of the Galaxy promise to revolutionize the field of Galactic archeology. We make predictions about how future \textit{Gaia} errors and the inclusion of [Fe/H] and age information will affect measurements of clustering in chemodynamical space. We predict that gains in {\it Gaia} precision will increase $\xi(\Delta r)$ by a factor of 1.5, a metallicity uncertainty of $\sigma_{[Fe/H]} = 0.05$ dex will increase the TPCF by a factor of two, and even $50\%$ uncertain ages will significantly enhance the TPCF.

\end{itemize}

We expect that the discrepancies between the simulated Galaxy and {\it Gaia} data will lead to new insights regarding the clustered nature of star formation and non-axisymmetric, time-dependent components of the Galactic potential.

\acknowledgements

We thank Angus Beane, Anthony Brown, Lehman Garrison, Yan-Fei Jiang, Diederik Kruijssen, Hans-Walter Rix, and members of the Conroy group at Harvard for useful discussions and helpful comments. HMK  acknowledges  support  from  the  DOE  CSGF  under  grant  number DE-FG02-97ER25308. CC acknowledges support from the Packard Foundation. YST is supported  by the NASA Hubble Fellowship grant HST-HF2-51425.001 awarded by the Space Telescope Science Institute. The computations in this paper were run on the Odyssey cluster supported by the FAS Division of Science, Research Computing Group at Harvard University. 

This work has made use of data from the European Space Agency mission \textit{Gaia} (\url{https://www.cosmos.esa.int/gaia}), processed by the Gaia Data Processing and Analysis Consortium (DPAC, \url{https://www.cosmos.esa.int/web/gaia/dpac/consortium}). Funding for the DPAC has been provided by national institutions, in particular the institutions participating in the Gaia Multilateral Agreement. The Sloan Digital Sky Survey IV is funded by the Alfred P.Sloan Foundation, the U.S. Department of Energy Office of Science, and the Participating Institutions and acknowledges support and resources from the Center for High-Performance Computing at the University of Utah. 

{\it Software:} \texttt{CorrFunc} \citep{sinha2018corrfunc, sinha2020corrfunc} \texttt{IPython} \citep{perez2007ipython}, \texttt{Cython} \citep{behnel2010cython}, \texttt{Astropy} \citep{2013A&A...558A..33A, 2018AJ....156..123A}, \texttt{NumPy} \citep{van2011numpy}, \texttt{SciPy} \citep{scipy}, \texttt{scikit-Learn} \citep{pedregosa2011scikit}, \texttt{Matplotlib} \citep{hunter2007matplotlib}.

\newpage

\bibliographystyle{aasjournal}

\bibliography{tpcf}



\appendix

\begin{figure}
 \includegraphics[width=84mm]{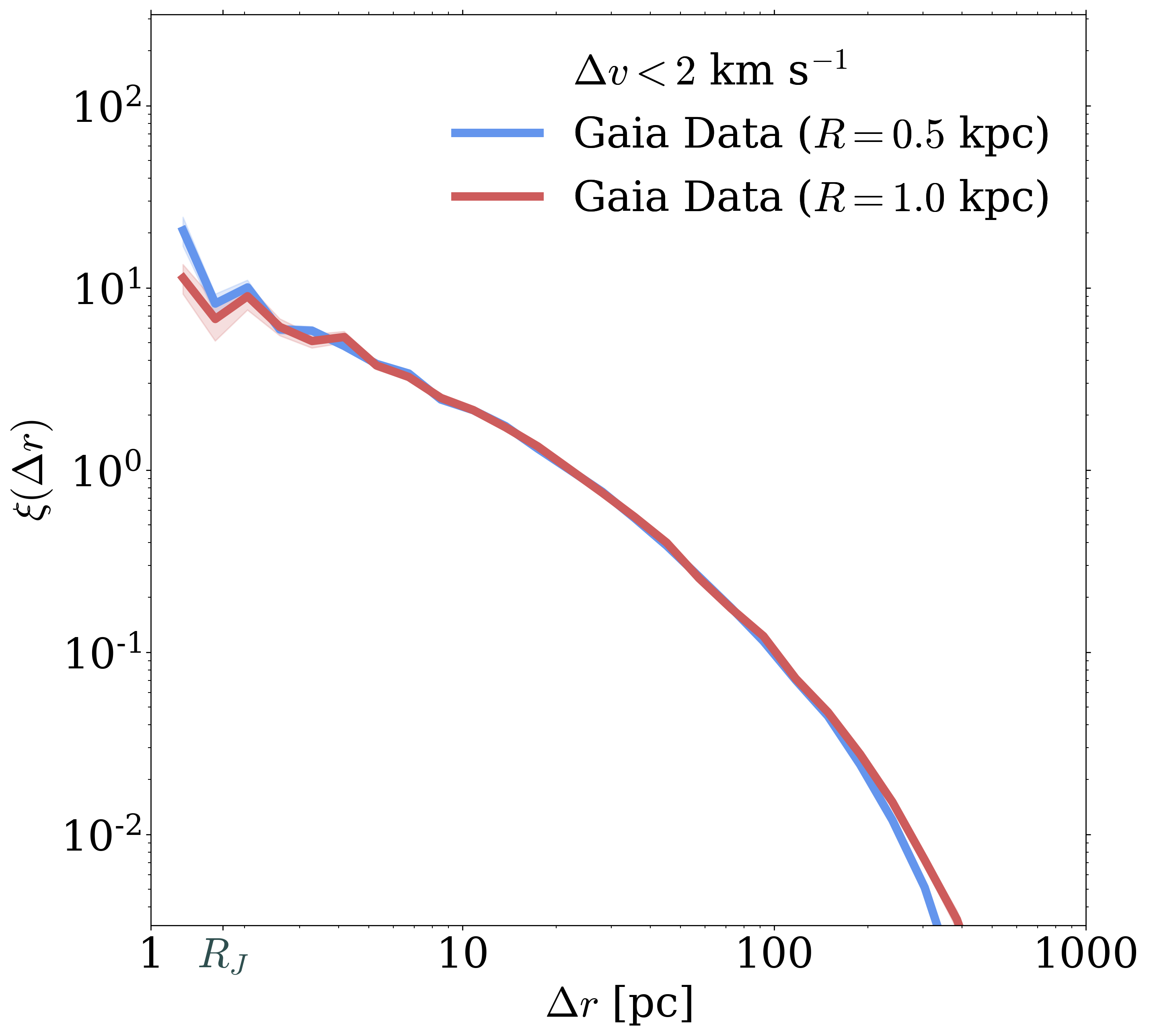}
 \caption{$\xi(\Delta r; \Delta v<2$ km s$^{-1})$ for solar cylinders of radii $R=0.5$ kpc and 1 kpc. The two TPCFs are nearly identical. We conclude that the rapid decline in the TPCF beyond $100$ pc is not a volume effect.}
 \label{fig:sp}
\end{figure} 

\begin{figure*}
 \includegraphics[width=168mm]{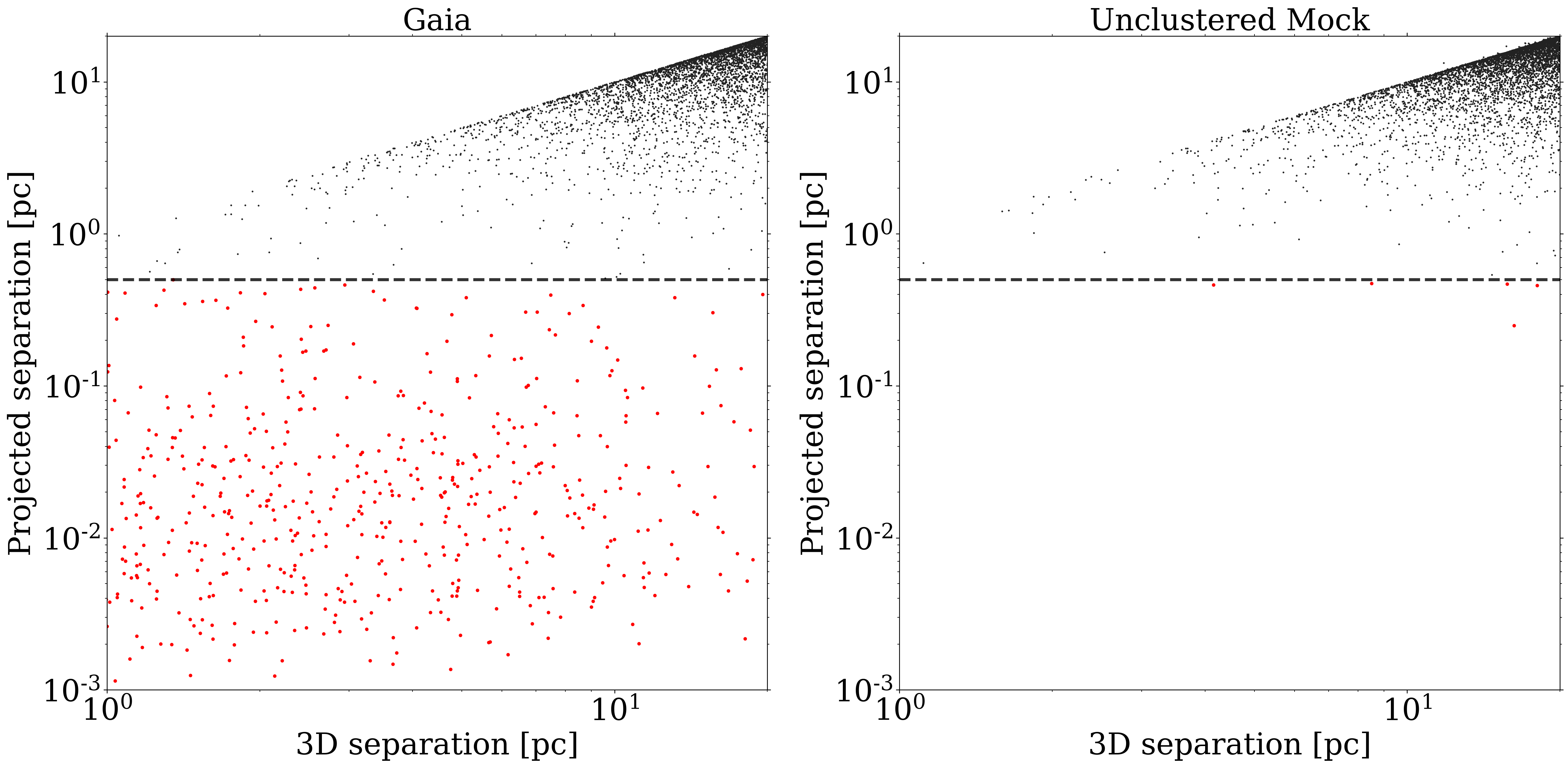}
 \caption{Projected separation and 3D separation of stars in \textit{Gaia} data (left panel) and an unclustered mock \citep{rybizki2018gaia} without wide binaries (right panel). Points with projected separation $< 0.5$ pc are colored red in both panels. There is a large population of pairs with very low projected separation relative to the 3D separation in the data but not in the unclustered mock. These pairs are likely bound wide binaries with true separations of $\ll 1 \rm pc$, whose calculated 3D separations were significantly inflated by parallax uncertainties. We only count pairs with projected separation $> 0.5$ pc in our TPCF calculations to minimize contamination from wide binaries.}
 \label{fig:wb_bound}
\end{figure*} 

\section{Volume Effects}
In our analyses we found a dramatic decline in the TPCF for both the data and the simulation at around $r \sim 200-400$ pc. Since the solar cylinders we consider here all have a radius of $0.5$ kpc, we test here whether or not the decline beyond $r \sim 200$ pc is due to volume effects. The volume we chose restricted stars to within $\sim 0.71$ kpc of the sun (largest possible distance from the sun with a cylinder radius of 0.5 kpc and height of 0.5 kpc above/below the sun). The TPCF considers any pair contained within the volume; the largest possible distance between any such pair of stars in this volume is $1.41$ kpc. However, there likely are not many pairs this far apart because most stars in this volume are contained within the thin disk (scale height $\sim 200$ pc). 

Figure \ref{fig:sp} shows the spatial TPCF of Gaia with the fiducial cylinder radius of $0.5$ kpc and a larger radius of $1$ kpc. The TPCFs for both volumes are almost identical. There are some discrepancies at the smallest spatial scales, which are driven by Poisson error and the fewer number of pairs. Overall, the excellent agreement confirms that the rapid decline in $\xi(\Delta r)$ at large $r$ is not due to our adopted geometry.

\section{The Impact of Wide Binaries on the TPCF}

\subsection{Bound Wide Binaries}

\begin{figure*}[!h]
 \includegraphics[width=168mm]{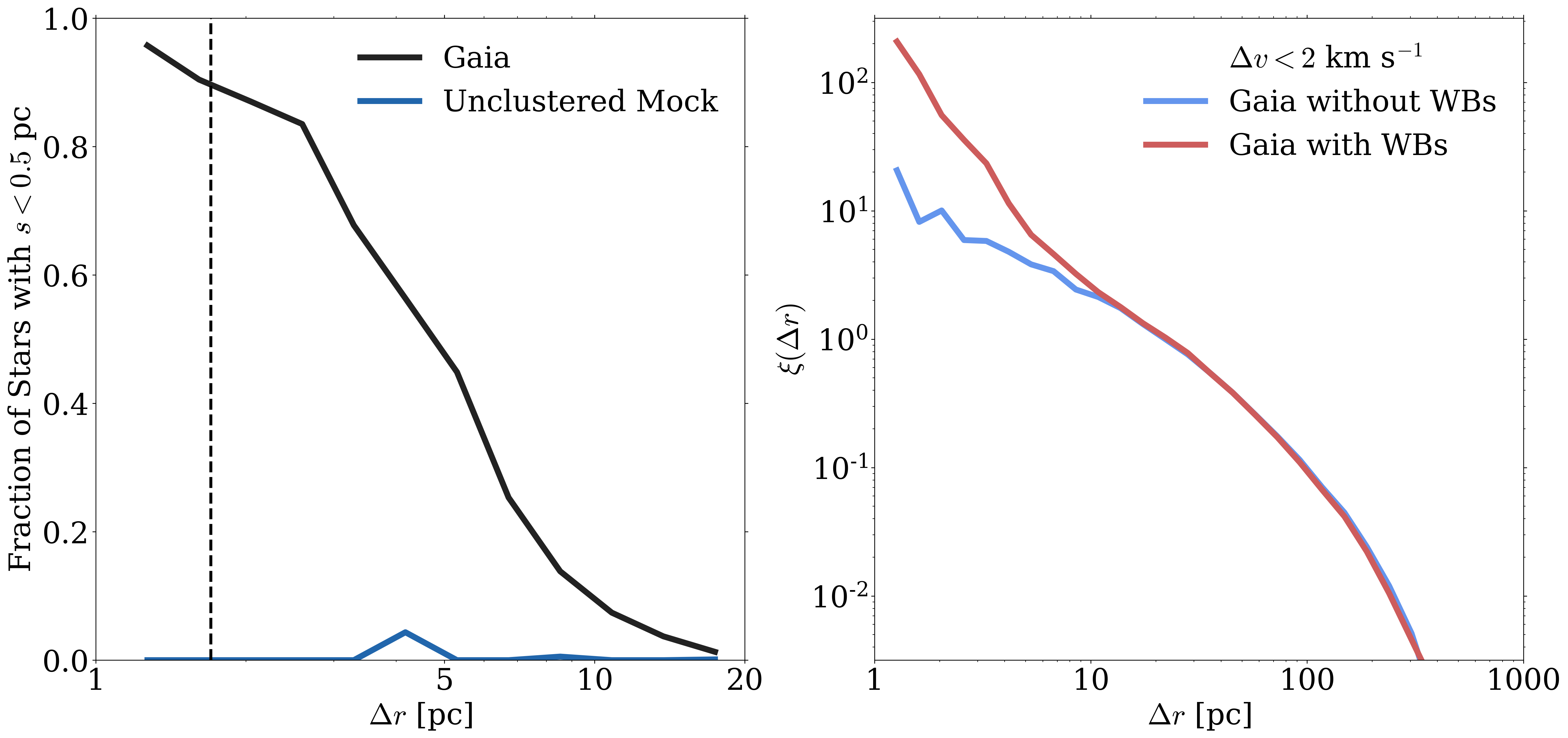}
 \caption{Fraction of pairs with projected separation $< 0.5$ pc in the \textit{Gaia} data and in the unclustered mock \citep{rybizki2018gaia}, and the corresponding change in the TPCF with and without the projected separation cut. \textit{Left panel:} The large fraction at low $\Delta r$ in the data, with no analog in the mock catalog, indicates contamination from bound wide binaries. This contamination is generally negligible for 3D separations $\Delta r \gtrsim 10\,\rm pc$. \textit{Right panel:} The spatial TPCF when pairs with projected separation less than $0.5$ pc are not counted. There is a large change in the clustering signal at $\Delta r < 20$ pc. }
 \label{fig:contam}
\end{figure*} 

Wide binaries are ubiquitous in the Galaxy \citep[e.g.,][]{jiang2010evolution, el2018wide, elbadry2019twins, tian2019separation}. Bound wide binaries are expected to have separations ranging from tens of AU to $\sim 1-2$ pc. The TPCF calculations presented in this work are all for pair separations of $>1$ pc. In the absence of observational uncertainties, we would thus not expect a significant fraction of pairs contributing to the TPCF to be bound binaries. However, observational uncertainties -- particularly the uncertainty in parallax -- make it difficult to measure the true 3D separation of close pairs. Our sample's median parallax uncertainty of $\sim 0.04$ mas corresponds to a distance uncertainty of $\pm 2$ pc at a distance of 200 pc, and $\pm 10$ pc at a distance of 500 pc. This means that parallax uncertainties can inflate the apparent separations of wide binaries that have true separations of $\ll 1 \rm$ pc up to $\sim 20 \rm pc$, and could lead to biases in the TPCF at $\Delta r \lesssim 20$\,pc if they they are not removed from the sample. 

``Stretching'' of wide binaries along the line of sight due to parallax uncertainties is a configuration space analog of the Fingers-of-God effect in redshift space. For binaries with small true separations, observational uncertainties significantly inflate their separations along the line of sight, but not their projected separations on the plane of the sky. 
We therefore calculate the projected separation of pairs (angular separation times the mean distance of the two stars) with $\Delta r < 20$ pc in the Gaia data and in the R18 mock (which contains no wide binaries). For true wide binaries, the projected separation should be much smaller than the calculated 3D separation. For non-binaries, the two separations should be comparable, since it is only for rare geometric alignments that the 3D separation is much greater than the 2D separation. The comparison between the 3D and the projected 2D separation is shown in Figure \ref{fig:wb_bound}. As expected, there is a large population of pairs with projected separation orders of magnitude smaller than the calculated 3D separation in the data, and almost none in the unclustered mock. 

The left panel of Figure \ref{fig:contam} shows the fraction of stars with projected separation $< 0.5$ pc for the data and the R18 mock. The contamination is $\sim 90\%$ below the Jacobi radius (dashed line), and falls to $\lesssim 10\%$ after 10 pc. The unclustered mock has very few pairs with projected separation $< 0.5$ pc out to 3D separations of $20$ pc. Consequently, our procedure of removing pairs with projected separations $< 0.5 \rm pc$ effectively selects bound wide binaries in the data, and removes very few non-binary pairs.  We therefore use this selection when computing the TPCF in the main paper. The right panel of Figure \ref{fig:contam} shows $\xi(\Delta r; \Delta v < 2$ km s$^{-1}$) with and without the projected separation cut. The clustering signals for $\Delta r \lesssim 10$ pc differ by an order of magnitude due to the presence of bound wide binaries. 

\subsection{Unbound Wide Binaries}

\begin{figure*}[!h]
 \includegraphics[width=168mm]{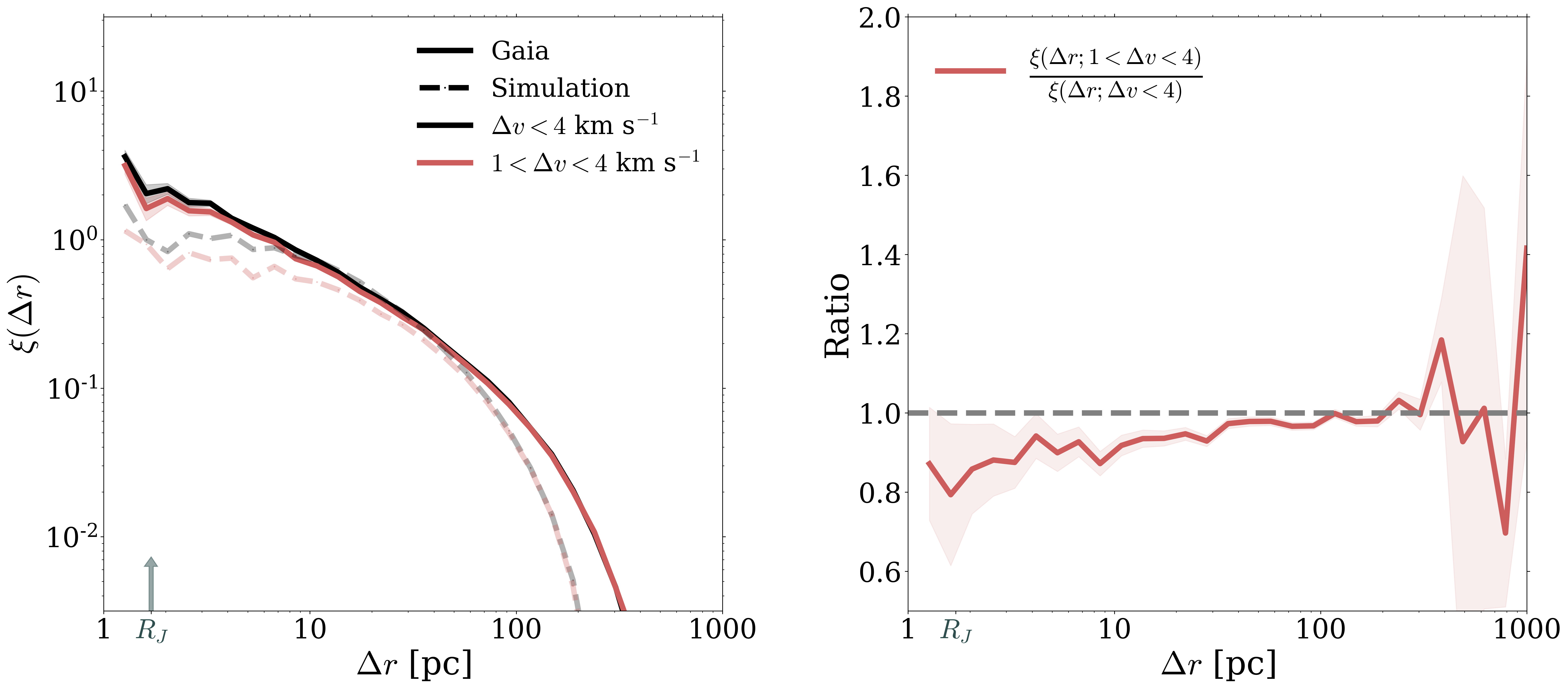}
 \caption{$\xi(\Delta r)$ for two different $\Delta v$ cuts to assess the contamination from unbound wide binaries. \textit{Left panel:} the spatial TPCF with these different velocity slices for the data and the simulation ($\Delta v < 4$ km s$^{-1}$, $1 < \Delta v < 4$ km s$^{-1}$).  Note that the simulation does not include wide binaries. \textit{Right Panel:} The ratio between $\xi(\Delta r; 1 < \Delta v < 4)$ and $\xi(\Delta r; \Delta v < 4)$. If unbound wide binaries were present in the data, we would expect them to be more common at smaller $\Delta v$.  The fact that the computed ratio is close to one suggests that unbound wide binaries are unlikely to affect the TPCF at a level beyond $\sim 5-10\%$ over the scales of interest.}
 \label{fig:wb}
\end{figure*} 

\cite{jiang2010evolution} (hereafter J10) simulated the orbital evolution and dissolution of wide binaries in the Galactic disk.  They argued that unbound wide binaries could remain close in phase space after being disrupted, which would lead to enhanced clustering at spatial separations extending beyond the Jacobi radius. Briefly, the models presented in J10 study the evolution of wide binaries due to gravitational perturbations from passing stars, and the Galactic tidal field. The component stars were tracked even after a wide binary become unbound. The discussion below considers the ``Opik 1'' model presented in J10, which  resembles the solar neighborhood. It is worth noting that the simulations presented in J10 do not include the impact of GMCs, which are likely an additional important scattering mechanism in the Galaxy \citep[e.g.,][]{weinberg1987dynamical}. 

We argued in K19b that there is a trough in the unbound wide binary separation distribution for the spatial scales $\sim 2 - 20$ pc. However, J10 predict that unbound wide binaries that are slowly drifting apart could lead to a peak in the separation distribution out to separations of $\sim 100-300$ pc for relative velocities $\Delta v \sim 0.1-0.2$ km s$^{-1}$. Almost all unbound wide binaries in the Opik 1 model have $\Delta v \leq 0.5$ km s$^{-1}$, with the majority at $\Delta v \leq 0.2$ km s$^{-1}$. However, given the uncertainties in the \textit{Gaia} data (especially the radial velocities), the computed $\Delta v$ for wide binaries could be a few times that. 

The contamination of wide binaries in the data can be estimated by contrasting the spatial TPCF for $\Delta v < 4$ with the TPCF for $1 < \Delta v < 4$ km s$^{-1}$. If many wide binaries are contributing to the TPCF signal, we would expect a precipitous drop-off in the TPCF when we probe the larger $1 < \Delta v < 4$ km s$^{-1}$ TPCF. The left panel of Figure \ref{fig:wb} shows the spatial TPCF with these different velocity slices for the data and the simulation. The simulation is plotted as a control to show what the same relative $\Delta v$ cut looks like in a mock without wide binaries; this is to motivate how much $\xi(\Delta r)$ changes just due to the change in the co-natal fraction. The right panel shows the fractions $\frac{\xi(\Delta r; 1 < \Delta v < 4)}{\xi(\Delta r; \Delta v < 4)}$ in red.

Both the data and the simulation show a fairly small change in $\xi(\Delta r)$ between $\Delta v < 4$ and $1 < \Delta v < 4$ km s$^{-1}$. $\xi(\Delta r)$ is smaller by a factor of $\sim 2$ at the smallest scales (perhaps due to bound wide binaries), and $\sim 1.1 - 1.3$ in the overdense region mentioned in J10 within the range $1 < \Delta v < 4$ compared to $\Delta v < 4$. A similar decrease in $\xi(\Delta r)$ is also seen in the fiducial simulation (left panel). These results indicate that wide binaries likely constitute a smaller overdensity in phase space at small $\Delta v$ and large $\Delta r$ than predicted in J10. 

\end{CJK*}
\end{document}